\documentclass{article}
\usepackage[latin1]{inputenc}   % Enter your text in ISO-Latin 1
\usepackage[dvips]{graphicx}
\usepackage{latexsym}
\usepackage{amssymb}
\usepackage{amsmath}

\makeatletter
\@ifclassloaded{siamltex}{} {%
\newtheorem{theorem}{Theorem}[section]
\newtheorem{lemma}[theorem]{Lemma}
\newtheorem{corollary}[theorem]{Corollary}
\newtheorem{proposition}[theorem]{Proposition}
\newtheorem{definition}[theorem]{Definition}
\newtheorem{example}[theorem]{Example}
\newtheorem{conjecture}[theorem]{Conjecture}
\newenvironment{keywords}{\textbf{Keywords.} }{}
\newenvironment{AMS}{\textbf{AMS:} }{}
\usepackage{ifthen}
\newenvironment{proof}[1][Default]
{\ifthenelse{\equal{Default}{#1}}{\noindent\textit{Proof. }}{\noindent\textit{Proof (#1). }}}
{\medskip}
\def\squareforqed{\hbox{\rlap{$\sqcap$}$\sqcup$}}
\def\qed{\ifmmode\squareforqed\else{\unskip\nobreak\hfil
\penalty50\hskip1em\null\nobreak\hfil\squareforqed
\parfillskip=0pt\finalhyphendemerits=0\endgraf}\fi}}
\makeatother

\author{Peter Jonsson\footnotemark[2] \and Fredrik Kuivinen\footnotemark[3] \and Gustav Nordh\footnotemark[4]}

\title{Approximability of Integer Programming with Generalised Constraints\footnotemark[1]}

\newcommand{\pr}[1]{\mathrm{Pr}\left[#1\right]}
\newcommand{\expect}[1]{\mathrm{E}\left[#1\right]}
\newcommand{\prob}[1]{{\sc #1}}
\newcommand{\cc}[1]{\textnormal{\textbf{#1}}} % Complexity Class
\newcommand{\opt}[0]{\textrm{{\sc opt}}}
\def\tup#1{\mathchoice{\mbox{\boldmath$\displaystyle#1$}}
{\mbox{\boldmath$\textstyle#1$}}
{\mbox{\boldmath$\scriptstyle#1$}}
{\mbox{\boldmath$\scriptscriptstyle#1$}}}

\begin{document}
\maketitle

\footnotetext[1]{Preliminary versions of parts of this article appeared in {\em Proceedings of the 12th International Conference on Principles and Practice of Constraint Programming}, Nantes, France, 2006}

%\footnotetext[2]{Department of Computer and Information Science, Linköpings Universitet, SE-581 83 Linköping, Sweden, $\{${\tt petej, freku, gusno}$\}${\tt @ida.liu.se}}

\footnotetext[2]{Department of Computer and Information Science, Linköpings Universitet, SE-581 83 Linköping, Sweden, email: {\tt petej@ida.liu.se}, phone: +46 13 282415, fax: +46 13 284499}

\footnotetext[3]{Department of Computer and Information Science, Linköpings Universitet, SE-581 83 Linköping, Sweden, email: {\tt freku@ida.liu.se}, phone: +46 13 286607, fax: +46 13 284499}

\footnotetext[4]{LIX, École Polytechnique, 91128 Palaiseau, France, email: {\tt nordh@lix.polytechnique.fr}, phone: +33 650479734}

\begin{abstract}
  We study a family of problems, called \prob{Maximum Solution}, where
  the objective is to maximise a linear goal function over the
  feasible integer assignments to a set of variables subject to a set
  of constraints.  When the domain is Boolean (i.e. restricted to
  $\{0,1\}$), the maximum solution problem is identical to the
  well-studied \prob{Max Ones} problem, and the approximability is
  completely understood for all restrictions on the underlying
  constraints [Khanna et al., SIAM J. Comput., 30 (2001),
  pp. 1863-1920].
  We continue this line of research by considering domains containing
  more than two elements.  We present two main results: a complete
  classification for the approximability of all maximal constraint
  languages over domains of cardinality at most $4$, and a complete
  classification of the approximability of the problem when the set of
  allowed constraints contains all permutation constraints. Under the
  assumption that a conjecture due to Szczepara holds, we give a
  complete classification for all maximal constraint languages. These
  classes of languages are well-studied in universal algebra and
  computer science; they have, for instance, been considered in
  connection with machine learning and constraint satisfaction.  Our
  results are proved by using algebraic results from clone theory and
  the results indicates that this approach is very powerful for
  classifying the approximability of certain optimisation problems.
\end{abstract}
  
\begin{keywords}combinatorial optimisation, constraint satisfaction, approximability, universal algebra\end{keywords}

\begin{AMS}08A70, 90C27, 68Q17, 68Q25\end{AMS}

\pagestyle{myheadings}
\thispagestyle{plain}
\markboth{GENERALISED MAX ONES}{P. JONSSON, F. KUIVINEN, AND G. NORDH}

\section{Introduction}

Our starting-point is the general combinatorial optimisation
problem {\sc Max Ones}$(\Gamma)$ where $\Gamma$ (known as the {\em constraint language}) 
is
a finite set of finitary relations
over $\{0,1\}$.  An instance of this problem consists of constraints
from $\Gamma$ applied to a number of Boolean variables, and the goal is to
find an assignment that satisfies all constraints while maximising the number
of variables set to 1.
It is easy to see that by choosing the constraint language 
appropriately, {\sc Max Ones}$(\Gamma)$
captures a number of well-known problems, for instance, 
{\sc Max Independent Set} (problem GT23 in~\cite{Kannetal99}),
and certain variants of {\sc Max 0/1 Programming} (problem MP2 in~\cite{Kannetal99}). Many other problems
are equivalent to {\sc Max Ones} under different reductions:
for instance, {\sc Max Set Packing} (also known as {\sc Max Hypergraph Matching})
and
{\sc Max Ones} are equivalent under {\sc Ptas}-reductions~\cite{Ausiello:etal:jcss80}.
%We also note that {\sc Max Ones} has been used for defining complexity
%classes (the $W$-hierarchy) in parameterised complexity~\cite{Downey:Fellows:PC}.

The approximability (and thus the computational complexity) is known
for all choices of $\Gamma$~\cite{KSTW01}.
For any Boolean constraint language $\Gamma$,
{\scshape Max Ones}$(\Gamma)$ is either in \cc{PO} or is \cc{APX}-complete or
\cc{poly-APX}-complete or finding a solution of non-zero value is \cc{NP}-hard
or finding any solution is \cc{NP}-hard.  The exact borderlines between the
different cases are given in~\cite{KSTW01}.
Actually, two different problems are studied in \cite{KSTW01}: the
{\em weighted} problem (where each variable is assigned a non-negative weight
and the objective is to find a solution of maximum total weight), and the
{\em unweighted} problem (where each variable is assigned the
weight 1). They prove that
the approximability for the weighted and unweighted versions of the
problem coincides.

We will study a generalisation of {\sc Max Ones} where
variable domains are different from $\{0,1\}$: this allows us to capture
more problems than with {\sc Max Ones}. For instance, this enables the study
of certain problems in integer linear programming~\cite{HN94},
problems in multiple-valued logic~\cite{Jonsson:Nordh:mfcs2006},
and in equation solving over Abelian groups~\cite{K05}.
For larger domains, it seems significantly harder to
obtain an exact characterisation of approximability than in the
Boolean case. Such a characterisation would, for instance,
show whether the dichotomy conjecture for
constraint satisfaction problems is true or not -- a famous open question
which is believed to be difficult~\cite{csp-mmsnp}.
Hence, we exhibit restricted (but still fairly general) 
families of constraint languages where the approximability can be
determined.

Let us now formally define the problem that we will study: Let $D \subset
\mathbb{N}$ ({\em the domain}) be a finite set.  The set of all
$n$-tuples of elements from $D$ is denoted by $D^n$. Any subset of
$D^n$ is called an $n$-ary relation on $D$. The set of all finitary
relations over $D$ is denoted by $R_D$. A {\em constraint language} over a
finite set, $D$, is a finite set $\Gamma \subseteq R_D$.  Constraint
languages are the way in which we specify restrictions on our
problems.  The constraint satisfaction problem over the constraint
language $\Gamma$, denoted {\scshape Csp}$(\Gamma)$, is defined to be
the decision problem with instance $(V,D,C)$, where
\begin{itemize}
\item $V$ is a set of variables,
\item $D$ is a finite set of values (sometimes called a domain), and
\item $C$ is a set of constraints $\{C_1,\dots,C_q\}$, in which each
  constraint $C_i$ is a pair $(s_i,R_i)$ where $s_i$ is a list of
  variables of length $m_i$, called the constraint scope, and
  $R_i$ is an $m_i$-ary relation over the set $D$, belonging to
  $\Gamma$, called the constraint relation.
\end{itemize}
The question is whether there exists a solution to $(V,D,C)$ or not,
that is, a function from $V$ to $D$ such that, for each constraint in
$C$, the image of the constraint scope is a member of the constraint
relation.
To exemplify this definition, let $NAE$ be the following ternary
relation on $\{0,1\}$: $NAE = \{0,1\}^3 \setminus
\{(0,0,0),(1,1,1)\}$.  It is easy to see that the well-known
\cc{NP}-complete problem {\scshape Not-All-Equal Sat} can be expressed
as {\scshape Csp}$(\{NAE\})$.

The optimisation problem that we are going to study, \prob{Weighted Maximum
  Solution}$(\Gamma)$ (which we abbreviate {\scshape W-Max Sol}$(\Gamma)$) is
defined as follows:
  \begin{description}
  \item[Instance:] Tuple $(V,D,C,w)$, where $D$ is a finite
    subset of $\mathbb{N}$, $(V,D,C)$ is a {\sc Csp}($\Gamma$) instance, 
    and $w : V \rightarrow \mathbb{N}$ is a weight function.
  
  \item[Solution:] An assignment $f : V \rightarrow D$ to the
    variables such that all constraints are satisfied.

  \item[Measure:] $\sum\limits_{v \in V} w(v) \cdot f(v)$
  \end{description}

%We remark in passing that non-trivial exponential-time algorithms for
%{\scshape W-Max Sol} have recently been presented in~\cite{A05}.
%The problem {\scshape W-Max Sol} should not be confused with the {\sc
%  Max Csp} problem where the objective is to maximise the number of
%satisfied constraints. 

\begin{example}
Consider the domain $D=\{0,1\}$ and the binary relation $R=\{(0,0),(1,0),(0,1)\}$.
Then, {\scshape W-Max Sol}$(\{R\})$ is exactly the weighted {\sc Maximum Independent Set}
problem.
\end{example}

Although the {\scshape W-Max Sol}$(\Gamma)$ problem is only defined for finite constraint languages, we 
will, in order to simplify the presentation, sometimes deal with sets of relations which are infinite. 
For a (possible infinite)
set of relations $X$ we will say that {\scshape W-Max Sol}$(X)$ is tractable
if {\scshape W-Max Sol}$(Y)$ is tractable for every finite subset $Y$ of $X$. Here
``tractable'' may be containment in one of \cc{PO}, \cc{APX}, or \cc{poly-APX}.
Similarly, we say that {\scshape W-Max Sol}$(X)$ is hard if there is a finite
subset $Y$ of $X$ such that {\scshape W-Max Sol}$(Y)$ is hard. Here ``hard'' will be
one of \cc{APX}-hard, \cc{poly-APX}-hard or that it is \cc{NP}-hard to find feasible
solutions.

Note that our choice of measure function in the definition of {\scshape W-Max Sol}$(\Gamma)$ is just one
of several reasonable choices. Another reasonable alternative, used in~\cite{K05}, would be to let the domain $D$ be any finite set and introduce an additional function $g: D \rightarrow \mathbb{N}$ mapping elements from the domain to natural numbers.
The measure could then be defined as %$\sum\limits_{v \in V} w(v) \cdot g(f(v))$. 
$\sum_{v \in V} w(v) \cdot g(f(v))$. 
This would result in a parameterised problem {\scshape W-Max Sol}$(\Gamma, g)$ where the goal is to classify the
complexity of {\scshape W-Max Sol}$(\Gamma, g)$ for all combinations of constraint languages 
$\Gamma$ and functions $g$. Note that our definition of {\scshape W-Max Sol}$(\Gamma)$ is equivalent to
the definition of {\scshape W-Max Sol}$(\Gamma, g)$ if in addition $g$ is required to be injective.
One of our motivations for the choice of measure function in the definition of
{\scshape W-Max Sol}$(\Gamma)$ is to stay closer to the definition of integer programming.

Only considering finite constraint language is in many cases not very
restrictive.
Consider for instance integer programming over the bounded domain $\{0,\ldots,d-1\}$.
Each row in the constraint matrix can be viewed as an inequality
\[a_1x_1+a_2x_2+\ldots+a_kx_k \geq b.\]
Obviously, such an inequality is equivalent to the following three inequalities
\[\begin{array}{c}a_1x_1+a_2x_2+\ldots+a_{\lfloor k/2 \rfloor}x_{\lfloor k/2 \rfloor}-z \geq 0\\
-a_1x_1-a_2x_2-\ldots-a_{\lfloor k/2 \rfloor}x_{\lfloor k/2 \rfloor}+z \geq 0\\
z+a_{\lfloor k/2 \rfloor+1}+\ldots+a_kx_k \geq b \end{array}\]
where $z$ denotes a fresh variable that is given the weight 0 in the objective function.
By repeating this process, one ends up with a set
of inequalities where each inequality contains at most three variables, and the
optimal solution to this instance have the same measure as the original instance. 
There are at most
$2^d+2^{d^2}+2^{d^3}$ different inequalities of length $\leq 3$ since the domain
contains $d$ elements, that is, we have reduced the problem to one with a finite constraint language.
Finally, this reduction is polynomial-time: each inequality of length
$k$ in the original instance give rise to at most $3^{\lceil \log_2 k \rceil} = O(k^2)$ inequalities
and at most $O(k^2)$ new variables.

While the approximability of {\scshape W-Max Sol} is well-understood
for the Boolean domain, this is not the case for larger domains. For
larger domains we are aware of three results, the first one is a tight
(in)approximability results for {\scshape W-Max Sol}$(\Gamma)$ when
$\Gamma$ is the set of relations that can be expressed as linear
equations over $\mathbb{Z}_p$~\cite{K05} (see also \S\ref{sec:affine-apx-hard} where we
define the problem formally). The second result is due to
Hochbaum and Naor~\cite{HN94} and they study integer programming with
monotone constraints, i.e., every constraint is of the form $ax - by
\leq c$, where $x$ and $y$ are variables and $a, b \in \mathbb{N}$ and
$c \in \mathbb{Z}$. In our setting, their result is a polynomial time
algorithm for certain constraint languages. 
The third result is a study of the approximability of certain logically
defined constraint languages~\cite{Jonsson:Nordh:mfcs2006}.
%It is worth pointing out
%that restricting integer programming to totally unimodular matrices
%do not give tractability results (in our sense) for \prob{W-Max Sol}. This is because
%the unimodularity restriction cannot
%be captured by only restricting the constraint language.
The main goal of this article is to gain a better understanding 
of non-Boolean
{\scshape W-Max Sol} --- for doing so, we will adapt
the algebraic approach for {\scshape Csp}s~\cite{FinAlg,JACM} for studying
the approximability of {\scshape W-Max Sol}.

When the algebraic approach is applicable to a certain problem, there is an
equivalence relation on the constraint languages such that two
constraint languages which are equivalent under this relation 
have the same complexity. More specifically, two constraint languages are
in the same equivalence class if they generate the same \emph{relational clone}.
The relational clone generated by $\Gamma$, captures the expressive power of $\Gamma$ and 
is denoted by $\langle \Gamma \rangle$.
%Those equivalence
%classes are called \emph{relational clones}. 
Hence, instead of studying every
possible finite set of relations it is enough to study the relational clones.
%Relational clones can be defined via a closure operator 
%$\langle \cdot \rangle$ on constraint languages: 
Thus, given two
constraint languages $\Gamma_1$ and $\Gamma_2$ such that
$\langle \Gamma_1 \rangle=\langle \Gamma_2 \rangle$ then,
{\scshape W-Max Sol}$(\Gamma_1)$ and
 {\scshape W-Max Sol}$(\Gamma_2)$ are equivalent under
polynomial-time reductions.

The clone-theoretic approach for studying the complexity of {\scshape Csp}s has been
very successful: it has, for instance, made it possible to design new
efficient algorithms and to clarify the borderline between
tractability and intractability in many important cases.  In
particular the complexity of the {\scshape Csp} problem over three element
domains is now completely understood~\cite{CSP3}.
In addition to the {\scshape Csp} problem it is possible to use the tools from
universal algebra to prove complexity results in many other {\scshape Csp}-like
problems. One example of such a problem is the quantified constraint
satisfaction problem ({\scshape QCsp}), where variables can not only be existentially
quantified but also universally quantified. The complexity of {\scshape QCsp} has successfully been
attacked with the clone-theoretic
approach~\cite{qcsp-surj,qcsp-3el}.  Furthermore, the
\#{\scshape Csp} problem~\cite{CCSPconj} (where the number of solutions to a
{\scshape Csp} is counted) 
%and the logical circumscription
%problem~\cite{KirousisK03,KirousisK04} 
have also benefitted from this 
approach. However, it seems that this
technique cannot be used for some other {\scshape Csp}-like problems: notable
exceptions are {\scshape Max Csp}~\cite{JKK06} and the problem of
enumerating all solutions to a {\scshape Csp} instance~\cite{enum-csp}. For some
problems it is the case that the relational clones are a useable tool in the
boolean domain but not in larger domains. The enumeration problem is
one such case~\cite{enum-csp}.

We begin by proving that the algebraic approach is applicable
to {\scshape W-Max Sol} and this result can be found in Theorem~\ref{relclone}\footnote{The proof is easy to adapt to other problems such as {\scshape W-Min Sol}
(the minimisation version of {\scshape W-Max Sol}) and {\scshape AW-Max Sol} (where both positive and negative weights are allowed).}. 
In fact, we show that
given two
constraint languages $\Gamma_1$ and $\Gamma_2$ such that
$\langle \Gamma_1 \rangle=\langle \Gamma_2 \rangle$, then
{\scshape W-Max Sol}$(\Gamma_1)$ $S$-reduces to
 {\scshape W-Max Sol}$(\Gamma_2)$, and vice-versa.
An $S$-{\em reduction} is a certain strong approximation-preserving
reduction:
if $\langle \Gamma_1 \rangle=\langle \Gamma_2 \rangle$, then
$\Gamma_1$ and $\Gamma_2$ are very similar with respect to
approximability.
For instance, if {\scshape W-Max Sol}$(\Gamma_1)$ is
{\bf NP}-hard to approximate within some constant $c$, then
{\scshape W-Max Sol}$(\Gamma_2)$ is
{\bf NP}-hard to approximate within $c$, too.
The proof is accompanied by an example of how the
approach can be used ``in practice'' for proving
approximability results.
We note that the clone-theoretic approach was not used in the original
classification of {\scshape Max Ones} and, consequently, the techniques we use
differs substantially from those used in~\cite{KSTW01}.
The results that we prove with the aid of Theorem~\ref{relclone} are the following:

\medskip

\noindent
\underline{{\bf Result 1.}}
Our first result concerns the complexity of \prob{W-Max Sol} for
\emph{maximal} constraint languages.
A constraint
language $\Gamma$ is maximal if, for any $R \not\in \langle \Gamma \rangle$, $\Gamma
\cup \{R\}$ has the ability to express (in a sense to be formally
defined later on) every relation in $R_D$. Such languages have
attracted much attention lately: for instance, the 
complexity of the corresponding 
{\sc Csp} problems has been completely classified.
In~\cite{BKJ01} the complexity was classified for domains $|D| \leq 3$
and necessary conditions for tractability was proved for the general
case. More recently, in~\cite{Bulatov04}, it was proved that those
necessary conditions also are sufficient for tractability.
Maximal constraint languages have also been studied in the context of
machine learning~\cite{Dalmau:Jeavons:tcs2003} and
quantified {\sc Csp}s~\cite{Chen:stacs2005}, and they attract 
a great deal of attention in
universal algebra, cf. the
survey by Quackenbush~\cite{Quackenbush:am1995}.

Our results show that if $\Gamma$ is maximal and $|D| \leq 4$, then
{\scshape W-Max Sol}$(\Gamma)$ is either tractable, \cc{APX}-complete,
\cc{poly-APX}-complete, that finding any solution with non-zero
measure is \cc{NP}-hard, or that {\scshape Csp}$(\Gamma)$ is not
tractable. Moreover, we prove that under a conjecture by
Szczepara~\cite{S96} our classification of maximal constraint
languages extends to arbitrary finite domains.
In the conference version~\cite{Jonsson:etal:cp2006} of this article
we claimed that we had characterised the complexity of \prob{Max Sol}
for all maximal constraint languages. Unfortunately, there was a flaw
in one of the proofs. We have managed to repair some of it by proving
the weaker results as stated above, but the general case, when $|D| >
4$ and Szczepara's conjecture is not assumed to hold, remains open.
We also note that the different cases can be efficiently 
recognised, i.e. 
the approximability of a maximal constraint language $\Gamma$ can be
decided in polynomial time (in the size of $\Gamma$).

When proving this result, we identified a new large tractable class of
{\scshape W-Max Sol}$(\Gamma)$: {\em generalised
  max-closed} constraints.  This class (which may be of independent interest) 
significantly extend
some of the tractable classes of {\sc Max Ones} that were
identified by Khanna et al.~\cite{KSTW01}.
It is also related to {\em monotone} constraints
which have been studied in mathematical programming and 
computer science~\cite{Hochbaum:etal:mp93,HN94,W02}.
In fact, generalised max-closed constraints generalise monotone constraints
over finite domains.
A certain kind of generalised max-closed constraints are relevant
in constraint programming languages such as {\sc Chip}~\cite{vanHentenryck:etal:ai92} as is pointed
out in \cite{JC96}. It may thus be possible to extend such languages
with optimisation capabilities by using the techniques presented in this article.

\medskip

\noindent
\underline{{\bf Result 2.}}
We completely characterise the approximability of
{\scshape W-Max Sol}$(\Gamma)$
when $\Gamma$ contains all permutation constraints.  Such
languages are known as {\em homogeneous} languages and 
Dalmau~\cite{Dalmau05} has determined the complexity of
{\scshape Csp}$(\Gamma)$ for all such languages while
the complexity of the corresponding quantified
{\sc Csp}s has been studied by Börner et al.~\cite{Borner:etal:csl2003}.
Szendrei~\cite{Szendrei} provides a compact presentation
of algebraic results on homogeneous algebras and languages.

We show that {\scshape W-Max Sol}$(\Gamma)$
is either tractable, \cc{APX}-complete, \cc{poly-APX}-complete, or
that {\scshape Csp}$(\Gamma)$ is not tractable. The four different cases
can, just as in {\bf Result 1}, be efficiently recognised. 
The proof is based on 
the characterisation of homogeneous algebras by
Marczewski~\cite{M64} and
Marchenkov~\cite{M82}.
For each domain $D$, there exists a set of relations $Q_D$ such that
every tractable homogeneous constraint language on $D$ is a subset
of $Q_D$. The relations in $Q_D$ are invariant under a certain
operation $t:D^3 \rightarrow D$ (known as the {\em discriminator} on $D$)
and the algebra $(D;t)$ is an example of a {\em quasi-primal} algebra 
in the sense of Pixley~\cite{Pixley:mz70}.
We note that the tractable homogeneous constraint languages
have been considered
earlier in connection with {\em soft constraints}~\cite{Cohen:etal:ai2006}, 
i.e. constraints which allows different levels of `desirability' to
be associated with different value assignments~\cite{Bistarelli:etal:jacm97}.
In the terminology of \cite{Cohen:etal:ai2006}, these languages are invariant under
a $\langle {\sf Mjrty}_1,{\sf Mjrty}_2,{\sf Mnrty}_3 \rangle$ multimorphism.
We also note that the tractable homogeneous languages
extend the {\em width-2 affine} class
of {\sc Max Ones} that was
identified by Khanna et al.~\cite{KSTW01}.

We remark that we do not deal explicitly with the unweighted 
version of the problem (denoted {\scshape Max Sol}$(\Gamma)$), where all 
variables have weight 1. The reason for this is that the approximability
classifications for {\scshape Max Sol}$(\Gamma)$ can be deduced
from the classifications for {\scshape W-Max Sol}$(\Gamma)$ 
(for all constraint languages $\Gamma$ considered in this paper).
In fact, as we explain in Section~\ref{sec:concl}, 
{\scshape Max Sol}$(\Gamma)$ have the same approximability as
{\scshape W-Max Sol}$(\Gamma)$ when $\Gamma$ one of the constraint
languages considered in this article.

\medskip

The article is structured as follows: We begin by presenting some basics
on approximability in \S{\ref{secapprox}}. The algebraic approach
for studying {\scshape W-Max Sol} is presented in \S{\ref{secalgcsp}},
\S{\ref{sechard}} identifies certain hard constraint languages, and
\S{\ref{sectract}} contains some tractability results. We continue with
\S{\ref{secmax}} that contain
{\bf Result 1} and \S{\ref{secdich}} that contain
{\bf Result 2}. Finally, \S{\ref{sec:concl}} contains some final remarks.

\section{Approximability, Reductions, and Completeness}
\label{secapprox}
A {\em combinatorial optimisation problem} is defined over a set
of {\em instances} (admissible input data); each instance $I$ has
a finite set ${\sf sol}(I)$ of {\em feasible solutions}
associated with it. 
Given an instance $I$ and a feasible solution $s$ of $I$, $m(I,s)$ denotes
the positive integer {\em measure} of $s$. 
The objective is, given an
instance $I$, to find a feasible solution of {\em optimum} value with respect
to the measure $m$. 
The optimal value is the largest one for {\em maximisation}
problems and the smallest one for {\em minimisation} problems. A
combinatorial optimisation problem is said to be an \cc{NPO} problem if its instances and solutions can
be recognised in polynomial time, the solutions are
polynomially bounded in the input size, and the objective function
can be computed in polynomial time (see,
e.g.,~\cite{Kannetal99}).

We say that 
a solution $s\in {\sf sol}(I)$ to an instance $I$ of an
\cc{NPO} problem $\Pi$ is $r$-{\em approximate} if it is
satisfying
\[
\max{\left \{ \frac{m(I,s)}{\opt(I)},\frac{\opt(I)}{m(I,s)} \right \} }\le r,
\]
where $\opt(I)$ is the optimal value for a solution to $I$. An
approximation algorithm for an \cc{NPO} problem $\Pi$ has {\em
performance ratio} $\mathcal{R}(n)$ if, given any instance $I$ of
$\Pi$ with $|I|=n$, it outputs an $\mathcal{R}(n)$-approximate
solution.

We define
\cc{PO} to be the class of \cc{NPO} problems that can be solved (to
optimality) in polynomial time. An \cc{NPO} problem $\Pi$ is in
the class \cc{APX} if there is a polynomial-time approximation
algorithm for $\Pi$ whose performance ratio is bounded by a
constant.
Similarly, $\Pi$ is in
the class \cc{poly-APX} if there is a polynomial-time approximation
algorithm for $\Pi$ whose performance ratio is bounded by a
polynomial in the size of the input.
Completeness in \cc{APX} and \cc{poly-APX} is defined using appropriate reductions,
called $AP$-reductions and $A$-reductions respectively~\cite{CKS,KSTW01}.
$AP$-reductions are more sensitive than $A$-reductions and every $AP$-reduction
is also an $A$-reduction~\cite{KSTW01}. In this paper we will not need the added flexibility
of $A$-reductions for proving our \cc{poly-APX}-completeness results.
Hence, we only need the definition of $AP$-reductions.
\begin{definition}
An \cc{NPO} problem $\Pi_1$ is said to be {\em $AP$-reducible} to an
\cc{NPO} problem $\Pi_2$ if two polynomial-time computable functions
$F$ and $G$ and a constant $\alpha$ exist such that
\begin{itemize}
\item[(a)] for any instance $I$ of $\Pi_1$, $F(I)$ is an
instance of $\Pi_2$;

\item[(b)] for any instance $I$ of $\Pi_1$, and any feasible
solution $s'$ of $F(I)$, $G(I,s')$ is a feasible solution of
$I$;

\item[(c)] for any instance $I$ of $\Pi_1$, and any $r\ge 1$, if
$s'$ is an $r$-approximate solution of $F(I)$ then $G(I,s')$ is
an $(1+(r-1)\alpha+o(1))$-approximate solution of $I$ where the
$o$-notation is with respect to $|I|$.
\end{itemize}

An \cc{NPO} problem $\Pi$ is {\em \cc{APX}-hard} ({\em \cc{poly-APX}-hard}) if every problem in
\cc{APX} (\cc{poly-APX}) is $AP$-reducible ($A$-reducible) to it. If, in addition, $\Pi$ is in
\cc{APX} (\cc{poly-APX}),  then $\Pi$ is called {\em \cc{APX}-complete} ({\em \cc{poly-APX}-complete}).
\end{definition}
It is a well-known fact (see, e.g., \S{8.2.1}
in~\cite{Kannetal99}) that $AP$-reductions compose.  In some proofs we
will use another kind of reduction, $S$-reductions.  They are defined as
follows:
\begin{definition}
An \cc{NPO} problem $\Pi_1$ is said to be {\em $S$-reducible} to an
\cc{NPO} problem $\Pi_2$ if two polynomial-time computable functions
$F$ and $G$ exist such that
\begin{itemize}
\item[(a)] given any instance $I$ of $\Pi_1$, algorithm $F$
produces an instance $I'=F(I)$ of $\Pi_2$, such that the measure of
an optimal solution for $I'$, $\opt(I')$, is exactly
$\opt(I)$.

\item[(b)] given $I'=F(I)$, and any solution $s'$ to $I'$,
algorithm $G$ produces a solution $s$ to $I$ such that
$m_1(I,G(s'))=m_2(I',s')$, where $m_1$ is the measure for
  $\Pi_1$ and $m_2$ is the measure for $\Pi_2$.
\end{itemize}
\end{definition}
Obviously, the existence of an $S$-reduction from $\Pi_1$ to $\Pi_2$
imply the existence of an $AP$-reduction from $\Pi_1$ to $\Pi_2$.
The reason why we need $S$-reductions is that $AP$-reductions do not
(generally) preserve membership in \cc{PO}~\cite{KSTW01}.
We also note that $S$-reduction preserve approximation thresholds
exactly for problems in {\bf APX}: let $\Pi_1,\Pi_2$ be problems
in {\bf APX}, assume that it is {\bf NP}-hard to approximate $\Pi_1$
within $c$, and that there exists an $S$-reduction from $\Pi_1$ to $\Pi_2$.
Then, it is {\bf NP}-hard to approximate $\Pi_2$ within $c$, too.

In some of our hardness proofs, it will be convenient for us to use a
type of approximation-preserving reduction called
$L$-reduction~\cite{Kannetal99}.
\begin{definition}
An \cc{NPO} maximisation problem $\Pi_1$ is said to be {\em $L$-reducible} to an
\cc{NPO} maximisation problem $\Pi_2$ if two polynomial-time computable functions
$F$ and $G$ and positive constants $\beta$ and $\gamma$ exist such that
\begin{itemize}
\item[(a)] given any instance $I$ of $\Pi_1$, algorithm $F$
produces an instance $I'=F(I)$ of $\Pi_2$, such that the measure of
an optimal solution for $I'$, $\opt(I')$, is at most $\beta\cdot
\opt(I)$;

\item[(b)] given $I'=F(I)$, and any solution $s'$ to $I'$, algorithm
  $G$ produces a solution $s$ to $I$ such that $|m_1(I,s)-\opt(I)|\le
  \gamma\cdot |m_2(I',s')-\opt(I')|$, where $m_1$ is the measure for
  $\Pi_1$ and $m_2$ is the measure for $\Pi_2$.
\end{itemize}
\end{definition}
It is well-known (see, e.g., Lemma 8.2 in~\cite{Kannetal99}) that, if
$\Pi_1$ is $L$-reducible to $\Pi_2$ and $\Pi_1 \in \cc{APX}$ then
there is an $AP$-reduction from $\Pi_1$ to $\Pi_2$.

\section{Algebraic Approach}
\label{secalgcsp}
We sometimes need to define relations in terms of other relations, using certain logical formulas.
In these definitions we use the standard correspondence between constraints and relations:
a relation consists of all tuples of values satisfying the corresponding
constraint. Although, we sometimes use the same symbol for a constraint and its corresponding relation, the
meaning will always be clear from the context.
More specifically, for a relation $R$ with arity $a$ we will sometimes write $R(x_1,
\ldots, x_a)$ with the meaning $(x_1, \ldots, x_a) \in R$ 
and the constraint $((x_1, \ldots, x_a), R)$ will sometimes
be written as $R(x_1, \ldots, x_a)$. 

An {\em operation} on a finite set $D$ (the domain) is an arbitrary function
$f:D^k \rightarrow D$.  Any operation on $D$ can be extended in a
standard way to an operation on tuples over $D$ as follows: Let $f$
be a $k$-ary operation on $D$ and let $R$ be an $n$-ary relation over
$D$. For any collection of $k$ tuples, $\tup{t_1},\tup{t_2}, \dots,
\tup{t_k} \in R$, the $n$-tuple $f(\tup{t_1},\tup{t_2}, \dots
,\tup{t_k})$ is defined as follows: $f(\tup{t_1},\tup{t_2}, \dots,
\tup{t_k}) = (f(\tup{t_1}[1],\tup{t_2}[1], \dots, \tup{t_k}[1]),$
$f(\tup{t_1}[2],\tup{t_2}[2], \dots, \tup{t_k}[2]),\dots,$
$f(\tup{t_1}[n],\tup{t_2}[n], \dots, \tup{t_k}[n]))$, where
$\tup{t_j}[i]$ is the $i$-th component in tuple $\tup{t_j}$.  A
technique that has shown to be useful in determining the computational
complexity of {\scshape Csp}$(\Gamma)$ is that of investigating
whether the constraint language $\Gamma$ is invariant under certain families of
operations~\cite{JACM}.

  Now, let $R_i \in \Gamma$. If $f$ is an operation such that for all
  $\tup{t_1},\tup{t_2}, \dots, \tup{t_k} \in R_i$ $f(\tup{t_1},
  \tup{t_2}, \dots ,\tup{t_k}) \in R_i$, then $R_i$ is
  {\em invariant} (or, in other words, \emph{closed}) under $f$. If all constraint
  relations in $\Gamma$ are invariant under $f$, then $\Gamma$ is
  invariant under $f$. An operation $f$ such that $\Gamma$ is
  invariant under $f$ is called a {\em polymorphism} of $\Gamma$. The set of
  all polymorphisms of $\Gamma$ is denoted $Pol(\Gamma)$. Given a set
  of operations $F$, the set of all relations that are invariant under
  all the operations in $F$ is denoted $Inv(F)$. Whenever there is
only one operation under consideration, we write $Inv(f)$ instead of
$Inv(\{f\})$.

We will need a number of operations in the sequel: an operation $f$ over $D$ is said to be 
\begin{itemize}
\item a {\em constant} operation if $f$ is unary and $f(a) = c$ for all $a
  \in D$ and some $c \in D$;
\item a {\em majority} operation if $f$ is ternary and $f(a,a,b)=f(a,b,a) =
  f(b,a,a) = a$ for all $a,b \in D$;
\item a {\em binary commutative idempotent} operation if $f$ is binary,
  $f(a,a) = a$ for all $a \in D$, and $f(a,b) = f(b,a)$ for all $a,b
  \in D$;
\item an {\em affine} operation if $f$ is ternary and $f(a, b, c) = a - b +
  c$ for all $a, b, c \in D$ where $+$ and $-$ are the binary
  operations of an Abelian group $(D,+,-)$.
\end{itemize}

\begin{example}
  Let $D = \{0, 1, 2\}$ and let $f$ be the majority operation on $D$
  where $f(a,b,c) = a$ if $a$, $b$ and $c$ are all distinct.
  Furthermore, let
  \[
  R = \{(0,0,1),(1,0,0),(2,1,1),(2,0,1),(1,0,1)\}.
  \]
  It is then easy to verify that for every triple of tuples, $\tup{x},
  \tup{y}, \tup{z} \in R$, we have $f(\tup{x}, \tup{y}, \tup{z}) \in
  R$. For example, if $\tup{x} = (0,0,1), \tup{y} = (2,1,1)$ and
  $\tup{z} = (1,0,1)$ then
  \begin{align}
  f(\tup{x}, \tup{y}, \tup{z}) = &\bigg(f(\tup{x}[1], \tup{y}[1], \tup{z}[1]), f(\tup{x}[2], \tup{y}[2], \tup{z}[2]), f(\tup{x}[3], \tup{y}[3], \tup{z}[3])\bigg) = \notag \\
  &\big(f(0,2,1), f(0,1,0), f(1,1,1)\big) = (0, 0, 1) \in R . \notag
  \end{align}
  We can conclude that $R$ is invariant under $f$ or, equivalently, that $f$ is a polymorphism of $R$.
\end{example}

We continue by defining a closure operation $\langle \cdot \rangle$ on
sets of relations: for any set $\Gamma \subseteq R_D$ the set $\langle
\Gamma \rangle$ consists of all relations that can be expressed using
relations from $\Gamma \cup \{ =_D \}$ ($=_D$ is the equality relation on
$D$), conjunction, and existential quantification.  Intuitively,
constraints using relations from $\langle \Gamma \rangle$ are exactly
those which can be simulated by constraints using relations from
$\Gamma$.  The sets of relations of the form $\langle \Gamma \rangle$
are referred to as {\em relational clones}.  An alternative
characterisation of relational clones is given in the following
theorem.
\begin{theorem}[\cite{PK79}]
For every set $\Gamma \subseteq R_D$, $\langle \Gamma \rangle = Inv(Pol(\Gamma))$.
\label{invpolth}
\end{theorem}

The following theorem states that when we are studying the approximability of {\scshape W-Max Sol}$(\Gamma)$, it is sufficient to
consider constraint languages that are relational clones.
\begin{theorem} \label{relclone}
  Let $\Gamma$ be a constraint language and $\Gamma' \subseteq
  \langle \Gamma \rangle$ finite. Then {\scshape W-Max Sol}$(\Gamma')$
  is $S$-reducible to {\scshape W-Max Sol}$(\Gamma)$.
\end{theorem}
\begin{proof}
  Consider an instance $I=(V,D,C,w)$ of {\scshape W-Max
    Sol}$(\Gamma')$. We transform $I$ into an instance $F(I) =
  (V',D,C',w')$ of {\scshape W-Max Sol}$(\Gamma)$.
  
  For every constraint $C = ((v_1, \dots, v_m),R)$ in $I$,
  $R$ can be represented as
\[
\exists_{v_{m+1}},\dots, \exists_{v_n} R_1(v_{11}, \dots, v_{1n_1}) \land
\dots \land R_k(v_{k1}, \dots, v_{kn_k})
\]
where $R_1, \dots, R_k \in \Gamma \cup \{=_D\}$,
$v_{m+1},\dots, v_n$ are fresh variables, and $v_{11}, \dots
,v_{1n_1},$ $v_{21}, \dots , v_{kn_k} \in \{v_1, \dots, v_n\}$.
Replace the constraint $C$ with the constraints 
\[((v_{11} \dots, v_{1n_1}), R_1),\dots, ((v_{k1},\dots, v_{kn_k}), R_k),\]
add $v_{m+1},\dots,v_n$ to $V$, and extend $w$ so that $v_{m+1}, \dots
v_n$ are given weight $0$.
%This can always be done and in polynomial time since c_1 and c_2 ... c_n \in Gamma
If we repeat the same reduction for every constraint in $C$, then it results
in an equivalent instance %$S''= (V'',D,C'',w'')$ 
of {\scshape W-Max Sol}$(\Gamma_1 \cup \{=_D\})$.

For each equality constraint $((v_i, v_j), =_D)$,
we do the following:
\begin{itemize}
\item replace all
  occurrences of $v_j$ with $v_i$, update $w'$ so that the weight of
  $v_j$ is added to the weight of $v_i$, remove $v_j$ from $V$, and
  remove the weight corresponding to $v_j$ from $w'$; and
\item remove $((v_i, v_j), =_D)$ from $C$.
\end{itemize}
The resulting instance $F(I) = (V',D,C',w')$ of {\scshape W-Max
  Sol}$(\Gamma)$ has the same optimum as $I$ (i.e., $\opt(I) =
\opt(F(I))$) and has been obtained in polynomial time.

Now, given a feasible solution $S'$ for $F(I)$, let $G(I,S')$ be the
feasible solution for $I$ where:
\begin{itemize}
\item The variables in $I$ assigned by $S'$ inherit their value from
  $S'$.
\item The variables in $I$ which are still unassigned all occur in
  equality constraints and their values can be found by simply
  propagating the values of the variables which have already been
  assigned.
\end{itemize} 

It should be clear that $m(I, G(I,S')) = m(F(I), S')$ for any feasible
solution $S'$ for $F(I)$. Hence, the functions $F$ and $G$, as
described above, is an $S$-reduction from {\scshape W-Max
  Sol}$(\Gamma')$ to {\scshape W-Max Sol}$(\Gamma)$.  
\end{proof}

To exemplify the use of the results in this section, we prove the following tight
approximability result:

\begin{lemma}
  Let $\Gamma$ be a finite constraint language over the domain
  $\{0,1\}$. If \prob{Max Ones$(\Gamma)$} is in \cc{APX} and not in
  \cc{PO}, then there is a polynomial time approximation algorithm for
  \prob{Max Ones$(\Gamma)$} with performance ratio $2$, and it is
  \cc{NP}-hard to approximate \prob{Max Ones$(\Gamma)$} within $2 -
  \epsilon$, for any $\epsilon > 0$.
\end{lemma}
\begin{proof}[Sketch]
  It follows from the classification results in \cite{KSTW01} that if \prob{Max Ones$(\Gamma)$} is
  in \cc{APX} and not in \cc{PO}, then $\Gamma$ is closed under the
  affine function $f(x,y,z) = x-y+z \; ({\rm mod} \; 2)$. It also
  follows from~\cite[Lemma 6.6]{KSTW01} that \prob{Max Ones$(\Gamma)$} is approximable within 2.
  
  In the Boolean domain, the structure of all relational clones is known. This
  classification was made by Emil Post in~\cite{post} and is often
  referred to as Post's lattice. 
 A gentle introduction to
  boolean relations and Post's lattice can be found
  in~\cite{signew1,signew2}.
  
  By Theorem~\ref{relclone}, it is enough to study the
  relational clones. By studying Post's lattice and the results for \prob{Max
    Ones} in~\cite{KSTW01}, one can conclude that there are three
  relational clones which are interesting in our case (i.e., there are three
  relational clones such that \prob{Max Ones$(\Gamma)$} is in \cc{APX} but not in
  \cc{PO}). Those relational clones are called $IL_0$, $IL_2$ and $IL_3$
  and can be defined as follows~\cite{signew2}:
  \begin{align}
    IL_0 &= \{ x_1 + \cdots + x_k = 0 \; ({\rm mod} \; 2) \mid k \in \mathbb{N} \}
\notag \\
    IL_2 &= \{ x_1 + \cdots + x_k = c \; ({\rm mod} \; 2) \mid k \in \mathbb{N}, \; c \in \{0,1\}  \} 
\notag \\
    IL_3 &= \{ x_1 + \cdots + x_k = c \; ({\rm mod} \; 2) \mid k  \textrm{ even, }  c \in \{0,1\} \}\notag
\end{align}
  We get the following inclusions from Post's lattice:
  $IL_0 \subset IL_2$ and $IL_3 \subset IL_2$.
  
  It is proved in~\cite{K05} that for a certain finite subset $\Gamma$
  of $IL_3$, \prob{Max Ones$(\Gamma)$} is \cc{NP}-hard to approximate
  within $2-\epsilon$ for all $\epsilon > 0$. As $IL_3 \subset IL_2$
  we get that \prob{Max Ones$(\Gamma)$} is \cc{NP}-hard to approximate
  within $2-\epsilon$ for all $\epsilon > 0$ if $\langle \Gamma
  \rangle = IL_2$.
  
  What remains to be done is to prove \cc{NP}-hardness for
  approximating \prob{Max Ones$(\Gamma)$} within $2-\epsilon$ if $\langle \Gamma \rangle =
  IL_0$. 
  We do this with a reduction from \prob{Max-E3-Lin-2} which is the
  following problem: given a set of equations over $\mathbb{Z}_2$ with
  exactly three variables per equation, satisfy as many equations as
  possible. It is proved in~\cite{optimalinapp} that it is
  \cc{NP}-hard to approximate \prob{Max-E3-Lin-2} within $2 -
  \epsilon$ for any $\epsilon > 0$.
  
  Let $I$ be an instance of \prob{Max-E3-Lin-2}. We will construct an
  instance $I'$ of \prob{Max Ones$(\Gamma)$} for a subset $\Gamma$ of
  $IL_0$. Given an equation $x_1+x_2+x_3 = 1 \; ({\rm mod} \; 2)$ in $I$ (we can
  assume that all equations have $1$ on the right hand
  side~\cite{optimalinapp}), we add the equation $x_1+x_2+x_3 = z$ (where
  $z$ is a fresh variable that only occurs in one equation) to
  $I'$.  Furthermore, we assign the weight zero to $x_1, x_2$ and
  $x_3$ and the weight one to $z$.  It is not hard to see that a
  solution with measure $m$ to $I$ can easily be transformed into a
  solution with measure $m$ for $I'$.  It is also the case that a
  solution of measure $m$ for $I'$ can be seen as a solution with
  measure $m$ for $I$.
\end{proof}

\section{Hardness and Membership Results}
\label{sechard}
In this section, we first prove
some general \cc{APX} and \cc{poly-APX} membership results for 
{\scshape W-Max Sol}$(\Gamma)$.
We also prove \cc{APX}-completeness and \cc{poly-APX}-completeness 
for some particular constraint languages. Most of our hardness results 
in subsequent sections are based on these results. 

We begin by making the following easy but interesting observation: we
know from the classification of {\scshape W-Max Sol}$(\Gamma)$ over
the Boolean domain $\{0,1\}$ that there exist many constraint
languages $\Gamma$ for which {\scshape W-Max Sol}$(\Gamma)$ is
\cc{poly-APX}-complete. However, if $0$ is not in the domain, then
there are no constraint languages $\Gamma$ such that {\scshape W-Max
  Sol}$(\Gamma)$ is \cc{poly-APX}-complete.
\begin{proposition} \label{prop:csp-no0-APX}
  If {\scshape Csp}$(\Gamma)$ is in \cc{P} and $0 \notin D$, then
  {\scshape W-Max Sol}$(\Gamma)$ is in \cc{APX}.
\label{propinapx}
\end{proposition}
\begin{proof}
  It is proved in~\cite{C04} that if {\scshape Csp}$(\Gamma)$ is in
  \cc{P}, then we can also find a solution in polynomial time. It
  should be clear that this solution is a
  $\frac{\max(D)}{\min(D)}$-approximate solution. Hence, we have a
  trivial approximation algorithm with performance ratio
  $\frac{\max(D)}{\min(D)}$. 
\end{proof}

Next, we present a general membership result for
{\scshape W-Max Sol}$(\Gamma)$. The proof is similar to the proof of
the corresponding result for the Boolean domain in~\cite[Lemma
6.2]{KSTW01} so we omit the proof.
\begin{lemma}
\label{lem-inpolyapx}
Let $\Gamma^c = \{\Gamma \cup \{\{(d_1)\}, \dots, \{(d_n)\}\}$, where
$D = \{d_1,\dots,d_n\}$ (i.e., $\Gamma^c$ is the constraint language
corresponding to $\Gamma$ where we can force variables to take any
given value in the domain).  If {\scshape Csp}$(\Gamma^c)$ is in \cc{P},
then {\scshape W-Max Sol}$(\Gamma)$ is in \cc{poly-APX}.
\end{lemma}

We continue by proving the
\cc{APX}-completeness of some constraint languages.

\begin{lemma} \label{lemmahardness}
  Let $R=\{(a,a),(a,b),(b,a)\}$ and $a,b \in D$ such that $0 < a < b$.
  Then, {\scshape W-Max Sol}$(\{R\})$ is \cc{APX}-complete.
\end{lemma}
\begin{proof} 
  Containment in \cc{APX} follows from Proposition~\ref{prop:csp-no0-APX}.
  To prove the hardness result we give an $L$-reduction (with parameters $\beta = 4b$ and $\gamma =
  \frac{1}{b-a}$) from the \cc{APX}-complete problem {\scshape Independent
    Set} restricted to degree $3$ graphs~\cite{AK00} to {\scshape
    Max Sol}$(\{R\})$.  Given an instance $I = (V,E)$ of {\scshape
    Independent Set} (restricted to graphs of degree at most $3$
  and containing no isolated vertices), let $F(I) = (V,D,C)$ be the
  instance of {\scshape Max Sol}$(\{R\})$ where, for each edge
  $(v_i,v_j) \in E$, we add the constraint $R(x_i,x_j)$ to $C$. For any
  feasible solution $S'$ for $F(I)$, let $G(I,S')$ be the solution for
  $I$ where all vertices corresponding to variables assigned $b$ in
  $S'$ form the independent set.  We have $|V|/4 \leq \opt(I)$ and
  $\opt(F(I)) \leq b|V|$ so $\opt(F(I)) \leq 4b\opt(I)$. Thus, $\beta
  = 4b$ is an appropriate parameter.
  
  Let $K$ be the number of variables being set to $b$ in an arbitrary
  solution $S'$ for $F(I)$. Then, 
  \begin{align}
    |\opt(I) - m(I, G(I, S'))| &= \opt(I) - K \quad \textrm{ and} \notag \\
    |\opt(F(I)) - m(F(I), S')| &= (b - a)(\opt(I) - K). \notag
  \end{align}
  Hence,
  \[
  |\opt(I) - m(I,G(I,S')| = \frac{1}{b-a}|\opt(F(I)) - m(F(I),S')|
  \]
  and $\gamma = \frac{1}{b-a}$ is an appropriate parameter.
  \end{proof}

The generic \cc{poly-APX}-complete constraint languages are presented in
the following lemma.

\begin{lemma} \label{lemmahardness2}
Let $R=\{(0,0),(0,b),(b,0)\}$ and $b \in D$ such that $0 < b$.
Then, {\scshape W-Max Sol}$(\{R\})$ is \cc{poly-APX}-complete.
\end{lemma}
\begin{proof}
  It is proved in~\cite[Lemma 6.15]{KSTW01} that for $Q =
  \{(0,0),(0,1),(1,0)\}$, it is the case that {\scshape W-Max
    Sol}$(\{Q\})$ is \cc{poly-APX}-complete.  To prove the
  \cc{poly-APX}-hardness we give an $AP$-reduction from {\scshape W-Max
    Sol}$(\{Q\})$ to {\scshape W-Max Sol}$(\{R\})$.  Given an instance
  $I$ of {\scshape W-Max Sol}$(\{Q\})$, let $F(I)$ be the instance of
  {\scshape W-Max Sol}$(\{R\})$ where all occurrences of $Q$ has been
  replaced by $R$. For any feasible solution $S'$ for $F(I)$, let
  $G(I,S')$ be the solution for $I$ where all variables assigned $b$
  in $S'$ are instead assigned $1$.  It should be clear that this is an
  $AP$-reduction, since if $S'$ is an $\alpha$-approximate solution to
  $F(I)$, then $G(I,S')$ is an $\alpha$-approximate solution for $I$.
  
  To see that {\scshape W-Max Sol}$(\{R\})$ is in \cc{poly-APX}, let
  $D = \{d_1, \ldots, d_n\}$ and note that $\Gamma^c = \{R,
  \{(d_1)\}, \ldots, \{(d_n)\}\}$ is invariant under the
  $\min$ function. As the $\min$ function is associative, commutative
  and idempotent, {\scshape Csp}$(\Gamma^c)$ is solvable in polynomial
  time~\cite{JACM}. Hence, {\scshape W-Max Sol}$(\{R\})$ is in
  \cc{poly-APX} due to Lemma~\ref{lem-inpolyapx}
\end{proof}

\section{Tractable Constraint Languages}
\label{sectract}

In this section, we 
present tractability results for two classes of constraint languages: 
{\em injective} constraint languages and {\em generalised max-closed}
constraint languages.
The tractability of injective constraints follows from
Cohen {\em et al.}~\cite[Sec. 4.4]{Cohen:etal:ai2006} but we present a simple proof for increased
readability. The tractability result for generalised max-closed
constraints is new and its proof constitutes the main part of this section.

These two classes can be seen as substantial and nontrivial 
generalisations of the tractable classes known for the corresponding 
{\scshape (Weighted) Max Ones} problem
over the Boolean domain. There are only three tractable classes of
 constraint languages over the Boolean domain, namely width-$2$ affine,
$1$-valid, and weakly positive \cite{KSTW01}. Width-$2$ affine 
constraint languages are examples of injective constraint languages 
and the classes of $1$-valid and weakly positive constraint languages 
are examples of generalised max-closed constraint languages.  
The monotone constraints which are, for instance,
studied by Hochbaum et al.~\cite{Hochbaum:etal:mp93,HN94}
(in relation with integer programming) and
Woeginger~\cite{W02} (in relation with constraint satisfaction)  
are also related to
generalised max-closed constraints. Hochbaum \& Naor~\cite{HN94}
show that monotone constraints can be characterised as those constraints
that are simultaneously invariant under
the $\max$ and $\min$ operators. Hence, monotone constraints are
also generalised max-closed constraints as long as the underlying
domain is finite.

\subsection{Injective relations} \label{injectivetract}

We begin by formally defining {\em injective relations}.

\begin{definition} \label{definj} \label{def-quasi-primal}
  A relation, $R \in R_D$, is called \emph{injective} if there exists
  a subset $D' \subseteq D$ and an injective function $\pi : D'
  \rightarrow D$ such that 
\[R = \{(x, \pi(x)) \mid x \in D'\}.\]
\end{definition}
It is important to note that the function $\pi$ is {\em not} assumed
to be total on $D$.
Let $I^D$ denote the set of all injective relations on the domain
$D$ and let $\Gamma_I^D = \langle I^D \rangle$.

\begin{example}
  Let $D = \{0, 1\}$ and let $R = \{(x, y) \mid x, y \in D, \; x + y
  \equiv 1 \; ({\rm mod} \; 2) \}$. The relation $R$ is injective
  because the function $f : D \rightarrow D$ defined as $f(0) = 1$ and
  $f(1) = 0$ is injective.  More generally, let $G = (D', +, -)$ be an
  arbitrary Abelian group and let $c \in D'$ be an arbitrary group
  element. It is easy to see that the relation $\{(x, y) \mid x, y \in
  D', \; x + y = c\}$ is injective.
\end{example}

  $R$ is an example of a relation which is invariant under an affine
  operation. Such relations have previously been studied in relation
  with the \prob{Max Ones} problem in~\cite{KSTW01,K05}. We will give some
  additional results for such constraints in \S\ref{sec:affine}.
  With the terminology used in~\cite{KSTW01,K05}, $R$ is said to be {\em width-$2$
  affine}. The relations which can be expressed as the set of solutions to an
  equation with two variables over an Abelian group are exactly the
  width-$2$ affine relations, so the injective
  relations are a superset of the width-$2$ affine relations.

To see that {\scshape W-Max Sol}$(\Gamma)$
is in \cc{PO} for every finite constraint language 
$\Gamma \subseteq \langle I^D \rangle$, it is sufficient to prove that
{\scshape W-Max Sol}$(I^D)$ is in \cc{PO} by Theorem~\ref{relclone}.
%(note that
%$I^D$ is a finite set).
Given an instance of {\scshape W-Max Sol}$(I^D)$, consider the graph having the 
variables as vertices and edges between the
vertices/variables occurring together in the same constraint. Each connected 
component of this graph represents an independent subproblem
that can be solved separately. If a value is assigned to a variable/vertex, 
all variables/vertices in the same component will be forced to take a value by 
propagating this assignment. Hence, each connected component has at most $|D|$ 
different solutions (that can be easily enumerated) and an optimal one can be found in polynomial time.

\subsection{Generalised Max-Closed Relations}

We begin by giving the basic definition:

\begin{definition}
  A constraint language $\Gamma$ over a domain $D \subset \mathbb{N}$
  is {\em generalised max-closed} if and only if there exists a binary
  operation $f \in Pol(\Gamma)$ such that for all $a,b \in D$, 
  %$f$ satisfies the following two conditions: 
  \begin{enumerate}
  \item
  if $a \neq b$ and $f(a,b) \leq \min(a,b)$, then
  $f(b,a) > \max(a,b)$; and
  \item
  %for all $a \in D$ it holds that
  $f(a,a) \geq a$.
  \end{enumerate}
\label{def-genmaxclosed}
\end{definition}

In the conference version of this article~\cite{Jonsson:etal:cp2006},
the definition of generalised max-closed constraint languages was
slightly more restrictive.  The following two examples will clarify
the definition above.
\begin{example}
  Assume that the domain $D$ is $\{0,1,2,3\}$.
  As an example of a generalised max-closed relation consider
  \[
  R = \{(0, 0), (1, 0), (0, 2), (1, 2)\} .
  \]
  $R$ is invariant under $\max$ and is therefore generalised
  max-closed since $\max$ satisfies the properties of Definition~\ref{def-genmaxclosed}.
  Now, consider the relation $Q$ defined as
  \[
  Q = \{(0,1), (1,0), (2,1), (2,2), (2,3)\} .
  \]
  $Q$ is not invariant under $\max$ because
  \[
  \max((0, 1), (1, 0)) = (\max(0, 1), \max(1, 0)) = (1, 1) \notin Q.
  \]
  Let the operation $\circ : D^2
  \rightarrow D$ be defined by the following 
Cayley table:\footnote{Note that we write $x \circ y$ instead of $\circ(x,y)$.}
\[ \begin{array}{c|cccc}
\circ & 0 & 1 & 2 & 3 \\ \hline
0     & 0 & 2 & 2 & 3 \\
1     & 2 & 1 & 2 & 2 \\
2     & 2 & 2 & 2 & 3 \\
3     & 3 & 2 & 3 & 3 \\
\end{array}\]
Now, it is easy to verify that $Inv(\circ)$ is a set of
  generalised max-closed relations and that $Q \in Inv(\circ)$.
\end{example}

\begin{example}
Consider the relations $R_1$ and $R_2$ defined as,
\[
R_1 = \{(1,1,1), (1,0,0), (0,0,1), (1,0,1)\}
\]
and $R_2 = R_1 \setminus \{(1,1,1)\}$. The relation $R_1$ is $1$-valid
because the all-1 tuple is in $R_1$, i.e., $(1,1,1)
\in R_1$. $R_2$, on the other hand, is not $1$-valid but is weakly
positive\footnote{A relation is weakly positive if it can be expressed 
as a CNF formula having at most one negated variable in each clause.} because it is invariant under $\max$. Note that both $R_1$
and $R_2$ are generalised max-closed since $R_1$ is invariant under
$f(x,y) = 1$ and $R_2$ is invariant under $f(x, y) = \max(x, y)$. It
is in fact the case that every weakly positive relation is invariant
under $\max$ (more is true in the Boolean domain: a relation is weakly
positive if and only if it is invariant under $\max$), so the $1$-valid and weakly positive relations are 
subsets of the generalised max-closed relations.
\end{example}

The tractability of generalised max-closed constraint languages crucially
depends on the following lemma.
\begin{lemma} \label{lem-max-tuple}
  If $\Gamma$ is generalised max-closed, then all relations
\[
R = \{(d_{11},d_{12},\dots,d_{1m}),\dots,(d_{t1},d_{t2},\dots,d_{tm})\}
\]
in $\Gamma$ have the property that the tuple
\[
\tup{t_{\max}} = (\max\{d_{11},\dots,d_{t1}\},\dots,\max\{d_{1m},\dots,d_{tm}\})
\]
is in $R$, too.
\end{lemma}
\begin{proof}
  Assume that there is a relation $R$ in $\Gamma$ such that the tuple
  \[\tup{t_{\max}} =
(\max\{d_{11},\dots,d_{t1}\},\dots,\max\{d_{1m},\dots,d_{tm}\})\]
  is not in $R$.  Define the distance between two tuples to be the
  number of coordinates where they disagree (i.e. the Hamming
  distance).  Let $\tup{a}$ be a tuple in $R$ with minimal distance
  from $\tup{t_{\max}}$ and let $I$ denote the set of coordinates where
$\tup{a}$ agrees
  with $\tup{t_{\max}}$. By the assumption that $\tup{t_{\max}}$ is
  not in $R$, we know that the distance between $\tup{a}$ and
  $\tup{t_{\max}}$ is at least $1$. Hence, without loss of generality, 
  assume that $\tup{a}[1] \neq \tup{t_{\max}}[1]$ and that $\tup{a}[1]$ is
  maximal for all tuples in $R$ agreeing with $\tup{t_{\max}}$ on 
  the coordinates in $I$. Let $\tup{b}$ be a tuple in $R$ such that
  $\tup{b}[1]=\tup{t_{\max}}[1]$.
  
  Since $\Gamma$ is generalised max-closed, there exists
  an operation $f \in Pol(\Gamma)$ such that for all $a,b \in D$ ($a \neq b$), it holds that 
  $f(a,b) > \max(a,b)$ whenever $f(b,a) \leq \min(a,b)$.  Furthermore, for all
  $a \in D$ it holds that $f(a,a) \geq a$.  
  Now consider the tuple $\tup{x^n}$ ($n = |D|$) defined as follows:
  $\tup{x^1} = f(\tup{a},\tup{b})$ and
  
  $$\tup{x^{i+1}} = \left\{ \begin{array}{ll}
f(\tup{x^i},\tup{a}) & \;\;\;\; \textrm{if $f(\tup{x^i}[1],\tup{a}[1]) > \tup{a}[1]$}, \\
f(\tup{a},\tup{x^i}) & \;\;\;\; \textrm{otherwise}. \end{array} \right.
$$
  %$$\tup{x^{i+1} = f(\tup{x^i},\tup{a}) if f(\tup{x^i}[1],\tup{a}[1]) > \tup{a}[1]$$
  %\[
  %\underbrace{f(f(f(\dots
  %  f(\tup{b},\tup{a}),\dots,\tup{a}),\tup{a}),\tup{a})}_{f \textrm{
  %    applied } |D| \textrm{ times}} = \tup{x^*}.
  %\]

  We begin by proving that $\tup{x^n}$ agrees with $\tup{a}$ on all
coordinates in $I$. Let $\tup{z}$ be an arbitrary tuple in $R$.
Note that for each $i \in I$ such that $\tup{z}[i] \neq \tup{a}[i]$, it is the case that
$f(\tup{a}[i], \tup{z}[i]) \leq \min(\tup{a}[i],\tup{z}[i])$ implies that 
$f(\tup{z}[i], \tup{a}[i]) > \max(\tup{a}[i],\tup{z}[i])$.
Hence, as $\tup{a}[i] = \tup{t_{\max}}[i]$, we cannot have that
$f(\tup{a}[i], \tup{z}[i]) \leq \min(\tup{a}[i], \tup{z}[i])$. So, for each $\tup{z} \in R$
and $i \in I$, we must have $f(\tup{a}[i], \tup{z}[i]) > \min(\tup{a}[i], \tup{z}[i])$ whenever
$\tup{a}[i] \neq \tup{z}[i]$. By an analogous argument, it follows that
for each $\tup{z} \in R$ and $i \in I$ we must have 
$f(\tup{z}[i], \tup{a}[i]) > \min(\tup{a}[i], \tup{z}[i])$
whenever $\tup{a}[i] \neq \tup{z}[i]$.

This together with the fact that $f(d,d) \geq d$, for all $d \in D$, and that
$\tup{a}$ agrees with $\tup{t_{\max}}$ on $I$ implies that $f(\tup{a}, \tup{x^n})$ agrees with
$\tup{a}$ on $I$.

%This follows from the fact that $f(\tup{a}[i], \tup{z}[i]) > \min(\tup{a}[i],\tup{z}[i])$ and $f(\tup{z}[i], \tup{a}[i]) > \min(\tup{a}[i],\tup{z}[i])$ for all tuples $\tup{z} \in R$, $i \in I$, and $\tup{z}[i] \neq \tup{a}[i]$ (since, $f(\tup{a}[i], \tup{z}[i]) \leq \min(\tup{a}[i],\tup{z}[i])$ implies that 
%$f(\tup{z}[i], \tup{a}[i]) > \max(\tup{a}[i],\tup{z}[i])$, contradicting the fact that $f \in Pol(\Gamma)$).

We now show that $\tup{x^n}[1] > \tup{a}[1]$.
This follows from essentially
the same argument as above. 
First note that $f(\tup{a}[1],\tup{b}[1]) = \tup{x^1}[1] > \tup{a}[1]$. 
If $f(\tup{a}[1],\tup{b}[1]) \leq \min(\tup{a}[1], \tup{b}[1])$, then $f(\tup{b}[1], \tup{a}[1]) > \tup{b}[1]$
which is not possible since $\tup{b}[1] = \tup{t_{\max}}[1]$.
Hence, we must have
$f(\tup{a}[1],\tup{b}[1]) = \tup{x^1}[1] > \min(\tup{a}[1], \tup{b}[1])$. 
%since $b[1]=t_{\max}[1]$ and, thus, $f(b[1],a[1]) \leq
%\max(a[1],b[1])$ and, by the definition of $f$, we have $f(a[1],b[1]) > min(a[1],b[1]) = a[1]$.
Now, by the definition of $\tup{x}^{i+1}$, it follows that if $\tup{x^i}[1] > \tup{a}[1]$, then $\tup{x}^{i+1}[1] > \min(\tup{x^i}[1],\tup{a}[1]) = \tup{a}[1]$ (just note that at least one of $f(\tup{x^i}[1],\tup{a}[1])$ and $f(\tup{a}[1],\tup{x^i}[1])$ is strictly larger than $\min(\tup{x^i}[1],\tup{a}[1]) = \tup{a}[1]$).
Hence, it follows by induction that $\tup{x^n}[1] > \tup{a}[1]$.

Thus, we have a contradiction with the fact that $a[1]$ is maximal for all tuples in $R$ agreeing with
$\tup{t_{\max}}$ on 
the coordinates in $I$. Hence, our assumption was wrong
and $\tup{t_{\max}}$ is in $R$.
\end{proof}

%Before we can prove the tractability of generalised max-closed
%constraint languages we need to introduce some terminology.
The algorithm for solving {\scshape W-Max Sol}$(\Gamma)$ when $\Gamma$ is generalised max-closed
is a simple consistency-based algorithm. 
The algorithm, which is based on pair-wise consistency, closely follows the algorithm
for {\scshape Csp}s over max-closed constraint languages from~\cite{JC96}.

We first need to introduce some terminology.
%the notion of pair-wise 
%consistent constraint satisfaction problems.
\begin{definition}
  Given a constraint $C_i=(s_i,R_i)$ and a (ordered) subset
  $s'_i$ of the variables in $s_i$ where $(i_1,i_2, \dots ,i_k)$ are
  the indices in $s_i$ of the elements in $s'_i$. The {\em projection}
  of $C_i$ onto the variables in $s'_i$ is denoted by $\pi_{s'_i} C_i$
  and defined as: $\pi_{s'_i} C_i = C'_i = (s'_i,R'_i)$ where
  $R'_i$ is the relation $\{(\tup{a}[i_1],\tup{a}[i_2],\dots,\tup{a}[i_k]) \; | \;
  \tup{a} \in R_i\}$.
\end{definition}

\begin{definition}
  For any pair of constraints $C_i = (s_i,R_i)$,
  $C_j=(s_j,R_j)$, the {\em join} of $C_i$ and $C_j$, denoted
  $C_i \Join C_j$, is the constraint on $s_i \cup s_j$ containing all
  tuples $\tup{t}$ such that $\pi_{s_i}\{ \tup{t} \} \in R_i$ and
  $\pi_{s_j}\{ \tup{t} \} \in R_j$.
\end{definition}

\begin{definition}[\cite{filtering89}]
  An instance of a constraint satisfaction problem $I = (V,D,C)$ is
  {\em pair-wise consistent} if and only if for any pair of
  constraints $C_i = (s_i,R_i)$, $C_j=(s_j,R_j)$ in $C$, it
  holds that the constraint resulting from projecting $C_i$ onto the
  variables in $s_i \cap s_j$ equals the constraint resulting from
  projecting $C_j$ onto the variables in $s_i \cap s_j$, i.e.,
  $\pi_{s_i \cap s_j} C_i = \pi_{s_i \cap s_j} C_j$.
\end{definition}

We are now ready to prove the tractability of generalised max-closed
constraint languages. 

%  The proof is a based on the proofs of Theorem 4.2 and Corollary 4.3
%  from~\cite{JC96}. The proofs of Theorem 4.2 and Corollary 4.3 from
% ~\cite{JC96} are concerned with max-closed constraints, but the only
%  property about max-closed constraints that is exploited is the fact
%  that $t_{max}$ is in $R$ for all max-closed relations $R$.
\begin{theorem} \label{thm-tractable-genmax}
  If \/ $\Gamma$ is generalised max-closed, then {\scshape W-Max
    Sol}$(\Gamma)$ is in \cc{PO}.
\end{theorem}
\begin{proof}
  Since $Inv(f)$ is a relational clone, constraints built over
  $Inv(f)$ are invariant under taking joins and projections~\cite[Lemma
  2.8]{TCS98} (i.e., the underlying relations are still invariant under
  $f$).  It is proved in~\cite{filtering89} that any set of
  constraints can be reduced to an equivalent set of pair-wise
  consistent constrains in polynomial time. Since the set of pair-wise
  consistent constraints can be obtained by repeated application of
  the join and projection operations, the underlying relations in the
  resulting constraints are still in $Inv(f)$.

  Hence, given an instance $I = (V,D,C,w)$ of \prob{W-Max
    Sol$(Inv(f))$}, we can assume that the constraints in $C$ are
  pair-wise consistent.  We prove that for pair-wise consistent
  $C$, either $C$ has a constraint with a constraint relation that do
  not contain any tuples (i.e., no assignment satisfies the
  constraint and there is no solution)
  or we can find the optimal solution in polynomial time.
  
  Assume that $C$ has no empty constraints.  For each variable $x_i$,
  let $d_i$ be the maximum value allowed for that variable by some
  constraint $C_j$ (where $x_i$ is in the constraint scope of $C_j$).
  We will prove that $(d_1,\dots,d_n)$ is an optimal solution to $I$.
  Obviously, if $(d_1,\dots,d_n)$ is a solution to $I$, then it is the
  optimal solution. Hence, it is sufficient to prove that
  $(d_1,\dots,d_n)$ is a solution to $I$.
  
  Assume, with the aim of reaching a contradiction, that
  $(d_1,\dots,d_n)$ is not a solution to $I$. Then, there exists a
  constraint $C_j$ in $C$ not satisfied by $(d_1,\dots,d_n)$. Since
  the constraint relation corresponding to $C_j$ is generalised
  max-closed, there exists a variable $x_i$ in the constraint scope of
  $C_j$ such that $C_j$ has no solution where $d_i$ is assigned to
  $x_i$.  Note that it is essential that $C_j$ is generalised
  max-closed to rule out the possibility that there exist two
  variables $x_i$ and $x_j$ in the constraint scope of $C_j$ such that
  $C_j$ has two solutions $t,u$ where $t(x_i)=d_i$ and $u(x_j) = d_j$,
  but $C_j$ has no solution $s$ where $s(x_i) = d_i$ and $s(x_j)=d_j$.
  We know that there exists a constraint $C_i$ in $C$ having $x_i$ in
  its constraint scope and $d_i$ an allowed value for $x_i$. This
  contradicts the fact that $C$ is pair-wise consistent. Thus,
  $(d_1,\dots,d_n)$ is a solution to $I$. 
\end{proof}

\section{Maximal Constraint Languages}
\label{secmax}
A {\em maximal constraint language} $\Gamma$ is a constraint language such
that $\langle \Gamma \rangle \subset R_D$, and if $R \notin \langle
\Gamma \rangle$, then $\langle \Gamma \cup \{R\} \rangle = R_D$. That
is, the maximal constraint languages are the largest constraint
languages that are not able to express all finitary relations over
$D$. This implies, among other things, that there exists an operation
$f$ such that $\langle \Gamma \rangle = Inv(f)$ whenever $\Gamma$ is 
a maximal constraint language~\cite{Rosenberg:86}. 
Relational clones 
$\langle \Gamma \rangle$ such that $\Gamma$ is a maximal constraint language
are called maximal relational clones.
%$\Gamma$ is a maximal constraint language.
The complexity of the
{\scshape Csp}$(\Gamma)$ problem for all maximal constraint languages on domains $|D| \leq 3$ was determined in~\cite{BKJ01}. Moreover, it was shown in~\cite{BKJ01} that the only case that remained to be classified in order to 
extend the classification to all maximal constraint languages over a finite domain was the case where $\langle \Gamma \rangle = Inv(f)$ for binary commutative idempotent operations $f$. These constraint languages were finally classified by Bulatov in~\cite{Bulatov04}.
\begin{theorem}[\cite{Bulatov04,BKJ01}]
  Let $\Gamma$ be a maximal constraint language on an arbitrary finite
  domain $D$.  Then, {\scshape Csp}$(\Gamma)$ is in \cc{P} if $\langle
  \Gamma \rangle = Inv(f)$ where $f$ is a constant operation, a
  majority operation, a binary commutative idempotent operation, or an
  affine operation.  Otherwise, {\scshape Csp}$(\Gamma)$ is
  \cc{NP}-complete.
\label{maxcspclass}
\end{theorem}

In this section, we classify the approximability of {\scshape W-Max
Sol}$(\Gamma)$ for all maximal constraint languages $\Gamma$ over $|D|
\leq 4$.  Moreover, we prove that the only cases that remain to be
classified, in order to extend the classification to all maximal
constraint languages over finite domains, are constraint languages
$\Gamma$ such that $\langle \Gamma \rangle$ is invariant under a
binary commutative idempotent operation. We also prove that if a
certain conjecture regarding minimal clones generated by binary
operations, due to Szczepara~\cite{S96}, holds, then our
classification can be extended to capture also these last cases.
\begin{theorem} \label{maxclass}
  Let $\Gamma$ be maximal constraint language on a finite domain $D$, with $|D| \leq 4$, and $\langle \Gamma \rangle = Inv(f)$. %be a maximal relational
%  clone on an arbitrary finite domain $D$.
\begin{enumerate}
\item If $\Gamma$ is generalised max-closed or an injective constraint
  language, then {\scshape W-Max Sol}$(\Gamma)$ is in
  \cc{PO};
  
\item else if $f$ is an affine operation, a constant operation
  different from the constant $0$ operation, or a binary commutative
  idempotent operation satisfying $f(0,b) > 0$ for all $b \in D
  \setminus \{0\}$ (assuming $0 \in D$); or if $0 \notin D$ and $f$ is
  a binary commutative idempotent operation or a majority operation,
  then {\scshape W-Max Sol}$(\Gamma)$ is \cc{APX}-complete;
  
\item else if $f$ is a binary commutative idempotent operation or a
  majority operation, then {\scshape W-Max Sol}$(\Gamma)$ is
  \cc{poly-APX}-complete;
  
\item else if $f$ is the constant $0$ operation, then finding a
  solution with non-zero measure is \cc{NP}-hard;
  
\item otherwise, finding a feasible solution is \cc{NP}-hard.
\end{enumerate}
Moreover, if Conjecture 131 from~\cite{S96} holds, then the results above hold for arbitrary finite domains $D$. 
\end{theorem}

The proof of the preceding theorem consists of a careful analysis of the approximability of 
{\scshape W-Max Sol}$(\Gamma)$ for all maximal constraint languages $\Gamma$ such that $\langle \Gamma \rangle = Inv(f)$, where $f$ is
one of the types of operations in Theorem~\ref{maxcspclass}. These results are
presented below.

\subsection{Constant Operation}
 We begin by considering maximal constraint languages that are
invariant under constant operations. Given an instance $I = (V,D,C)$ of a
\prob{Csp} problem, we define the \emph{constraint graph} of $I$ to be
$G = (V, E)$ where $\{v, v'\} \in E$ if there is at least one
constraint $c \in C$ which have both $v$ and $v'$ in its constraint
scope.

\begin{lemma}
\label{lemmaconstclosed}
Let $d^*=\max(D)$ and let $C_d$ be a constraint language such that
$\langle C_d \rangle = Inv(f_d)$ where $f_d:D \rightarrow D$
satisfies $f_d(x)=d$ for all $x \in D$. Then,
{\scshape W-Max Sol}$(C_{d^*})$ is in \cc{PO},
{\scshape W-Max Sol}$(C_{d})$ is \cc{APX}-complete if
$d \in D \backslash \{d^*,0\}$, and
it is \cc{NP}-hard to find a solution with non-zero measure for {\scshape W-Max Sol}$(C_{0})$.
\end{lemma}
\begin{proof}
  The tractability of {\scshape W-Max Sol}$(C_{d^*})$ is trivial,
  since the optimum solution is obtained by assigning $d^*$ to all
  variables.
  
  For the \cc{APX}-hardness of {\scshape W-Max Sol}$(C_{d})$ ($d \in D
  \backslash \{d^*,0\}$), it is sufficient to note that
  $\{(d,d),(d,d^*),(d^*,d)\}$ is in $\langle C_d \rangle $, and since $0<d<d^*$ it
  follows from Lemma~\ref{lemmahardness} that {\scshape W-Max
    Sol}$(C_{d})$ is \cc{APX}-hard. It is easy to realise that
  {\scshape W-Max Sol}$(C_{d})$ is in \cc{APX}, since we can obtain a
  $\frac{d^*}{d}$-approximate solution by assigning the value $d$ to
  all variables.
  
  The fact that it is \cc{NP}-hard to find a solution with non-zero
  measure for {\scshape W-Max Sol}$(C_{0})$ over the Boolean domain
  $\{0,1\}$ is proved in~\cite[Lemma 6.23]{KSTW01}. To prove that it
  is \cc{NP}-hard to find a solution with non-zero measure for
  {\scshape W-Max Sol}$(C_{0})$ over a domain $D$ of size $\geq 3$, we
  give a reduction from the well-known \cc{NP}-complete problem
  {\scshape Positive-1-in-3-Sat}~\cite{GJ79}, i.e., {\scshape
    Csp}$(\{R\})$ with $R = \{(1,0,0), (0,1,0), (0,0,1)\}$. 
    It is easy to see that {\scshape Positive-1-in-3-Sat} restricted to instances
    where the constraint graph is connected is still \cc{NP}-complete.
    
  Now, let $R' = \{(b,a,a),(a,b,a),(a,a,b),(0,0,0)\}$, where $0<a<b$
  and $a,b,0 \in D$. For an instance $I = (V,D,C)$ of {\scshape
    Csp}$(\{R\})$ where the constraint graph of $I$ is connected,
  create an instance $I'$ of {\scshape W-Max Sol}$(\{R'\})$ by
  replacing all occurrences of $R$ by $R'$ and giving all variables
  weight $1$.  Since the constraint graph is connected, 
$I$ has a solution if and only if $I'$ has a
  solution with non-zero measure, and since $R' \in C_0$, it follows
  that it is \cc{NP}-hard to find a solution with non-zero measure for
  {\scshape W-Max Sol}$(C_{0})$. 
\end{proof}

%\subsection{Majority Operation}
%Maximal constraint languages based on majority operations are
%fairly easy to analyse due to the results in \S~\ref{sechard}.
%\begin{lemma}
%\label{lemmamajclosed}
%Let $m$ be an arbitrary majority operation on $D$ and $\langle \Gamma_m \rangle = Inv(m)$ . Then, {\scshape
%  W-Max Sol}$(\Gamma_m)$ is \cc{APX}-complete if $0 \notin D$ and
%\cc{poly-APX}-complete if $0 \in D$.
%\end{lemma}
%\begin{proof}
%  Arbitrarily choose elements $a,b \in D$ such that $a < b$.  Then, it is easy to see that
%  $\{(a,a),(a,b),(b,a)\}$ is in $\langle \Gamma_m \rangle$. Thus, by
%  Proposition~\ref{propinapx} and Lemmas~\ref{lemmahardness}
%  and~\ref{lemmahardness2}, it follows that {\scshape W-Max
%    Sol}$(\Gamma_m)$ is \cc{APX}-complete or \cc{poly-APX}-complete
%  depending on whether $0$ is in $D$ or not. 
%\end{proof}

\subsection{Majority Operation}
Maximal constraint languages based on majority operations are
fairly easy to analyse due to the results in \S\ref{sechard}.
\begin{lemma}
\label{lemmamajclosed}
Let $m$ be an arbitrary majority operation on $D$. Then, {\scshape
  W-Max Sol}$(Inv(m))$ is \cc{APX}-complete if $0 \notin D$ and
\cc{poly-APX}-complete if $0 \in D$.
\end{lemma}
\begin{proof}
  Arbitrarily choose elements $a,b \in D$ such that $a < b$.  Then, it is easy to see that
  $\{(a,a),(a,b),(b,a)\}$ is in $Inv(m)$. Thus, by
  Proposition~\ref{propinapx} and Lemmas~\ref{lemmahardness}
  and~\ref{lemmahardness2}, it follows that {\scshape W-Max
    Sol}$(Inv(m))$ is \cc{APX}-complete or \cc{poly-APX}-complete
  depending on whether $0$ is in $D$ or not. 
\end{proof}

\subsection{Affine Operation} \label{sec:affine}

We split the proof of this result into two parts.
The first part, \S\ref{sec:affine-apx-hard}, contains the hardness result:
for every affine operation $a:D^3 \rightarrow D$, 
%there
%exists a finite subset $\Gamma$ of $Inv(a)$ such that
\prob{W-Max Sol}$(Inv(a))$ is \cc{APX}-hard.
The proof is based on a reduction from
\prob{Max-$p$-Cut} which is
a well-known \cc{APX}-complete problem~\cite{Kannetal99}. Membership in
\cc{APX} is proved in \S\ref{sec:affine-apx} by presenting an approximation
algorithm with constant performance ratio.

We will denote the affine operation on the group $G$ by $a_G$, i.e.,
if $G = (D, +_G, -_G)$ then $a_G(x, y, z) = x -_G y +_G z$.

\subsubsection{\cc{APX}-hardness} \label{sec:affine-apx-hard}
In this section we will prove Theorem~\ref{th:affine-apx-hard}, which
states that relations invariant under an affine operation give rise to
\cc{APX}-hard \prob{W-Max Sol}-problems. We need a number of lemmas
before we can prove this result. We begin
by giving an $L$-reduction from {\scshape Max-$p$-Cut} to {\scshape W-Max Sol
Eqn}$(\mathbb{Z}_p, g)$ where $p$ is prime. \prob{Max-$p$-Cut}
and \prob{W-Max Sol Eqn} are defined as follows:
\begin{definition}[\cite{Kannetal99}]
  \prob{Max-$p$-Cut} is an optimisation problem with
  \begin{description}
  \item[Instance:] A graph $G = (V, E)$.
  \item[Solution:] A partition of $V$ into $p$ disjoint sets $C_1,
    C_2, \ldots, C_p$.
  \item[Measure:] The number of edges between the disjoint sets, i.e.,
  \[
  \sum_{i=1}^{p-1} \sum_{j=i+1}^p |\{ \{v, v'\} \in E \mid v \in C_i \textrm{ and } v' \in C_j \}|.
  \]
  \end{description}
\end{definition}

\begin{definition}[\cite{K05}]
  Let $G = (D, +_G, -_G)$ be a group
  and $g : D \rightarrow \mathbb{N}$ a function. \prob{W-Max Sol Eqn$(G, g)$} is an optimisation
  problem with
  \begin{description}
  \item[Instance:] A triple $(V, E, w)$ where, $V = \{v_1, v_2,
    \ldots, v_n\}$ is a set of variables, $E$ is a set of equations of
    the form $u_1 +_G \ldots +_G u_k = 0_G$, where each $u_i$ is
    either a variable (e.g., ``$v_4$''), an inverted variable (e.g.,
    ``$-_G \: v_7$'') or a group constant, and $w$ is a weight function $w
    : V \rightarrow \mathbb{N}$.
    
  \item[Solution:] An assignment $f : V \rightarrow D$ to the
    variables such that all equations are satisfied.

  \item[Measure:] $\sum\limits_{v \in V} w(v)g(f(v))$
  \end{description}
\end{definition}

We do not require the group $G$ to be Abelian in the definition of
\prob{W-Max Sol Eqn} but this will always be the case in this article.
Note that the function $g$ and the group $G$ are not parts of
the input so \prob{W-Max Sol Eqn$(G, g)$} is a problem
parameterised by $G$ and $g$.  We refer the reader
to~\cite{K05} for more information on the problem
{\scshape W-Max Sol Eqn}$(\mathbb{Z}_p, g)$,

The following lemma follows from the proof of Proposition~2.3
in~\cite{supmod-maxcsp}.
\begin{lemma} \label{lem:pcut-opt}
  For any instance $I = (V, E)$ of {\scshape Max-$p$-Cut}, we have
  $\opt(I) \geq |E| (1 - 1/p)$.
\end{lemma}

We can now prove the \cc{APX}-hardness of \prob{W-Max Sol Eqn}.
\begin{lemma} \label{lemmaequationhardness} \label{lem:maxsoleqn-apx-hard}
  For every prime $p$ and every non-constant function $g :
  \mathbb{Z}_p \rightarrow \mathbb{N}$, {\scshape W-Max Sol
  Eqn}$(\mathbb{Z}_p, g)$ is \cc{APX}-hard.
\end{lemma}
\begin{proof}
  Given an instance $I = (V, E)$ of \prob{Max-$p$-Cut}, we construct
  an instance $F(I)$ of {\scshape W-Max Sol Eqn}$(\mathbb{Z}_p, g)$
  where, for every vertex $v_i \in V$, we create a variable $x_i$ and
  give it weight $0$, and for every edge $\{v_i, v_j\} \in E$, we
  create $p$ variables $z^{(k)}_{ij}$ for $k = 0, \ldots, p-1$ and
  give them weight $1$. Let $g_{\min}$ denote an element in
  $\mathbb{Z}_p$ that minimises $g$, i.e.,
  \[
  \min_{x \in \mathbb{Z}_p} g(x) = g(g_{\min})
  \]
  and let $g_s$ denote the sum
  \[
  \sum_{k = 0}^{p-1} g(k) .
  \]
  For every edge $\{v_i, v_j\} \in E$, we introduce the equations
  \[
  k(x_i - x_j) + g_{\min} = z^{(k)}_{ij}
  \]
  for $k = 0, \ldots, p-1$. If $x_i = x_j$, then the $p$
  equations for the edge $\{v_i, v_j\}$ will contribute $pg(g_{\min})$
  to the measure of the solution. On the other hand, if $x_i \neq x_j$
  then the $p$ equations will contribute $g_s$ to the measure.

  Given a solution $s'$ to $F(I)$, we can construct a solution $s$ to $I$
  in the following way: let $s(v_i) = s'(x_i)$, i.e. for every vertex
  $v_i$, place this vertex in partition $s'(x_i)$. The measures of the
  solutions $s$ and $s'$ are related to each other by the equality
  \begin{equation} \label{eq:soleqn-sol}
  m'(F(I), s') = |E| \cdot p \cdot g(g_{\min}) + (g_s - p \cdot g(g_{\min})) \cdot m(I, s).
  \end{equation}
  From \eqref{eq:soleqn-sol}, we get
  \begin{equation} \label{eq:soleqn-opt}
  \opt(F(I)) = |E| \cdot p \cdot g(g_{\min}) + (g_s - p \cdot g(g_{\min})) \cdot \opt(I) 
  \end{equation}
  and from Lemma~\ref{lem:pcut-opt}, we have that $\opt(I) \geq |E| \cdot (1 -
  1/p)$ which implies $\opt(I) \geq |E|/p$. By combining this with
  \eqref{eq:soleqn-opt}, we can conclude that
  \begin{align}
  \opt(F(I)) &=    \opt(I) \left( \frac{|E| \cdot p \cdot g(g_{\min})}{\opt(I)} + g_s - p \cdot g(g_{\min}) \right) \notag \\
            &\leq \opt(I) \Big( p^2 \cdot g(g_{\min}) + g_s - p \cdot g(g_{\min}) \Big) . \notag
  \end{align}
  Hence, $\beta = p(p - 1) \cdot g(g_{min}) + g_s$ is an appropriate
  parameter for the $L$-reduction.
  
  We will now deduce an appropriate $\gamma$-parameter for the
  $L$-reduction: from \eqref{eq:soleqn-sol} and \eqref{eq:soleqn-opt} we get
  \[
  |\opt(F(I)) - m'(F(I), s')| = (g_s - p \cdot g(g_{\min})) \cdot |\opt(I) - m(I,s)| 
  \]
  so, $\gamma = 1/(g_s - p \cdot g(g_{\min}))$ is sufficient ($\gamma$ is
  well-defined because a non-constant $g$ implies $g_s > p \cdot g_{\min}$).
\end{proof}

We need two lemmas before we can prove the \cc{APX}-hardness of affine
relations.  Let $v_1, v_2, \ldots, v_k$ be a collection of
variables, $G = (D, +_G, -_G)$ an Abelian group, and $E$
an equation of the form $x_1 +_G x_2 +_G \ldots +_G x_n = c$,
where each $x_i$ is a (possibly inverted) variable and $c
\in D$. Note that each variable may occur several times in $E$. The set of
all solutions to $E$ may be seen as a $k$-ary relation $R_E$ on $D^k$. The
following two lemmas are well-known~\cite{JACM}.
\begin{lemma} \label{lem:closedEq}
The relation $R_E$ is invariant under $a_G$.
\end{lemma}

\begin{lemma} \label{lem:closedCoset}
If $P$ is a coset of $G$, then $P$ is invariant under $a_G$.
\end{lemma}

We now have all results needed to prove the main theorem of
this section.
\begin{theorem} \label{th:affine-apx-hard}
  \prob{W-Max Sol$(Inv(a_G))$} is \cc{APX}-hard for every affine operation
  $a_G$.
\end{theorem}
\begin{proof}
  We show that there exists a prime $p$ and a non-constant function $h
    : \mathbb{Z}_p \rightarrow \mathbb{N}$ such that \prob{W-Max Sol
    Eqn$(\mathbb{Z}_p, h)$} can be $S$-reduced to \prob{W-Max
    Sol$(Inv(a_G))$}. The result will then follow from
    Lemma~\ref{lem:maxsoleqn-apx-hard}.
  
  Let $p$ be a prime such that $\mathbb{Z}_p$ is isomorphic to a
  subgroup $H$ of $G$. We know that such a $p$ always exists by the fundamental theorem of finitely
  generated Abelian groups.
  Let
  $\alpha$ be the isomorphism which maps elements of $\mathbb{Z}_p$ to
  elements of $H$ and let $h = \alpha$. (Note that $H \subset
  \mathbb{N}$ since the domain is a subset of
  $\mathbb{N}$. Consequently, $h$ may be viewed as a function from
  $\mathbb{Z}_p$ to $\mathbb{N}$.)

  Let $I = (V, E, w)$ be an instance of \prob{W-Max Sol
  Eqn$(\mathbb{Z}_p, h)$} with variables $V = \{v_1, \ldots, v_n\}$ and equations $E =
  \{e_1, \ldots, e_m\}$. We will construct an instance $I' =
  (V,D,C,w)$ of \prob{W-Max Sol$(Inv(a_G))$}.
  
  Let $U$ be the unary relation for which $x \in U \iff x \in H$; this
  relation is in $Inv(a_G)$ by Lemma~\ref{lem:closedCoset}.  For every
  equation $E_i \in E$, there is a corresponding pair $(s_i, R_i)$
  where $s_i$ is a list of variables and $R_i$ is a relation in
  $Inv(a_G)$ such that the set of solutions to $E_i$ are exactly the
  tuples which satisfies $(s_i, R_i)$ by Lemma~\ref{lem:closedEq}. We
  can now construct $C$:
  \[
  C = \{ (v_i, U) \ | \ 1 \leq i \leq n \} \cup \{ (s_i, R_i) \ \mid \ 1 \leq i \leq m \} .
  \]
  It is easy to see that $I$ and $I'$ are essentially the same in the
  sense that every feasible solution to $I$ is also a feasible
  solution to $I'$, and they have the same measure. The converse is
  also true: every feasible solution to $I'$ is also a feasible
  solution to $I$. Hence, we have given a $S$-reduction from
  \prob{W-Max Sol Eqn\mbox{$(\mathbb{Z}_p, h)$}} to \prob{W-Max
  Sol$(Inv(a_G))$}. As $h$ is not constant (it is in fact injective),
  it follows from Lemma~\ref{lem:maxsoleqn-apx-hard} that \prob{W-Max
  Sol Eqn\mbox{$(\mathbb{Z}_p, h)$}} is \cc{APX}-hard. This
  $S$-reduction implies that \prob{W-Max Sol$(Inv(a_G))$} is
  \cc{APX}-hard.
\end{proof}

\subsubsection{Membership in \cc{APX}} \label{sec:affine-apx}
We will now prove that relations that are invariant under an affine
operation give rise to problems which are in \cc{APX}.  It has been
proved that a relation which is invariant under an affine operation is
a coset of a subgroup of some Abelian group~\cite{JACM}.  We will give
an approximation algorithm for the more general problem when the
relations are cosets of subgroups of a finite group.

Our algorithm is based on an algorithm by Bulatov and
Dalmau~\cite{maltsev-simple} for deciding the satisfiability of
{\em Mal'tsev constraints}. A \emph{Mal'tsev operation} is a
ternary operation $m$ such that $m(x, y, y) = m(y, y, x) = x$ for all
$x, y \in D$. 
If a constraint language $\Gamma$ is invariant under a Mal'tsev operation,
then Bulatov and Dalmau have proved that {\sc Csp}$(\Gamma)$ is solvable in
polynomial time. We note that every affine operation is a Mal'tsev
operation since $x-_G y+_G y=x$ and $y-_G y+_G x = x$. 

Let $G^k$ denote the direct product of $k$ copies of $G$.  We are now
ready to prove containment in \cc{APX}.
\begin{theorem}
\label{thmaffinapx}   
  Let $G = (D; +_G, -_G)$ be a finite group and let $\Gamma$ be a
  constraint language such that for each $R \in \Gamma$ there is an
  integer $k$ such that $R$ is a coset of some subgroup of $G^k$. Then
  {\scshape W-Max Sol}$(\Gamma)$ is in \cc{APX}.
\end{theorem}
\begin{proof}
  Let $I = (V,D,C,w)$ be an arbitrary instance of
  {\scshape W-Max Sol}$(\Gamma)$
  where $V = \{v_1,\ldots, v_n\}$. Feasible solutions to our optimisation problem can
  be viewed as certain elements in $H = G^n$.  Each constraint $C_i \in
  C$ defines a coset $a_i +_G J_i$ of $H$ with representative $a_i
  \in H$, for some subgroup $J_i$ of $H$. The set of solutions to the
  problem is the intersection of all those cosets. Thus, $S =
  \bigcap_{i=1}^{|C|} a_i +_G J_i$ denotes the set of all solutions.

  Since $\Gamma$ is invariant under
  the affine operation
 $a_G(x, y, z)
  = x -_G y +_G z$ and $a_G$ is a Mal'tsev operation, we can
  decide if there are any solutions to $I$ in
  polynomial-time~\cite{maltsev-simple}.  Clearly, $S$ is empty if and only if
  there are no solutions. 
It is
  well-known that an intersection of a set of cosets is either empty
  or a coset so if $S \neq \emptyset$, then $S$ is a coset.

  We will represent the elements of $G^n$ by vectors $\tup{x} = (x_1,
  \ldots, x_n)$ where each $x_i$ is an element of $G$.  For any
  instance $I$, we define $\mathcal{R}(I)$ to be the random variable
  which is uniformly distributed over the set of solutions to $I$. Let
  $V_i$ denote the random variable which corresponds to the value
  which will be assigned to $v_i$ by $\mathcal{R}(I)$.  We claim that
  $V_i$ is uniformly distributed over some subset of $G$. As $S$ is a
  coset there is a subgroup $S'$ of $G^n$ and an element $\tup{s} \in
  S$ such that $S = \tup{s} +_G S'$. Assume, for the sake of
  contradiction, that $V_i$ is not uniformly distributed. Then, there
  are group elements $a, b \in G$ such that the sets
  \[
  X_a = \{ \tup{x} \in S' \mid x_i = a \} \textrm{ and } X_b = \{ \tup{x} \in S' \mid x_i = b \}
  \]
  have different cardinality. Assume that $|X_a| > |X_b|$. Arbitrarily
  pick $\tup{y} \in X_a, \tup{z} \in X_b$ and construct the set $Z =
  \{ \tup{x} -_G \tup{y} +_G \tup{z} \mid \tup{x} \in X_a
  \}$. From the definition of $Z$ and the fact that $S'$ is invariant
  under $a_G$, it follows that $Z \subseteq S'$. For each $\tup{x} \in
  Z$ we have $x_1 = b$, hence $Z \subseteq X_b$. However, we also have
  $|Z| = |X_a|$, which contradicts the assumption that $|X_a| >
  |X_b|$. We conclude that this cannot hold and $V_i$ is
  uniformly distributed. Hence, for each $1 \leq i \leq n$, $V_i$ is
  uniformly distributed.

  Now, let $A$ denote the set of indices such that for every $i \in
  A$, $\pr{V_i = c_i} = 1$ for some $c_i \in G$. That is, $A$ contains
  the indices of the variables $V_i$ which are constant in every
  feasible solution. Let $B$ contain the indices for the variables
  which are not constant in every solution, i.e., $B = [n] \
  \backslash \ A$.

Let $S^* = \sum_{i \in B} w(v_i) \max(D) + \sum_{i \in A} w(v_i) c_i$
and note that $S^* \geq \opt$. Furthermore, let
\[
E_{\min} = \min_{X \subseteq G, |X| > 1} \frac{1}{|X|} \cdot \sum_{x \in X} x
\]
and note that $\max(D) > E_{\min} > 0$.

The expected value of the measure of $\mathcal{R}(I)$ can now be
estimated as
\begin{align}
\expect{\sum_{i=1}^n w(v_i) V_i} &=    \sum_{i \in A} w(v_i) \expect{V_i} + 
\sum_{i \in B} w(v_i) \expect{V_i}  \label{eq:affineapx} \\
                                 &\geq \sum_{i \in A} w(v_i) c_i + E_{\min} 
\sum_{i \in B} w(v_i) \geq \frac{E_{\min}}{\max(D)} S^* \geq \frac{E_{\min}}{\max(D)} \opt . \notag
\end{align}
Since $E_{\min} / \max(D) > 0$, it follows that the measure of $\mathcal{R}(I)$ has, in
expectation, a constant performance ratio. We will denote
$\frac{E_{\min}}{\max(D)} \cdot \opt$ by $E$.

To get a deterministic polynomial-time algorithm, note that for any
instance $I$ we can use the algorithm by Bulatov and
Dalmau~\cite{maltsev-simple} to compute the two sums in~\eqref{eq:affineapx} in
polynomial-time. Hence, we can compute the expected measure of
$\mathcal{R}(I)$ in polynomial-time. Our algorithm is presented in
Figure~\ref{fig:alg-affineapx}.

\begin{figure}
\textbf{Input:} An instance $I = (V, D, C, w)$ of {\scshape W-Max
Sol}$(\Gamma)$ \\
\textbf{Output:} A solution with performance ratio at least $E_{\min}/\max(D)$, or ``no solution'' if there are no solutions.
\begin{enumerate}
  \item Return ``no solution'' if there are no solutions (use Bulatov and Dalmau's algorithm to check this)
  \item Let $I_1 = I$.
  \item For each $i$ from $1$ to $|V|$:
  \item $\qquad$ For each $x \in D$:
  \item $\qquad \qquad$ Let $I_i = I$ and add the constraint $v_i = x$ to $I_i$
  \item $\qquad \qquad$ If there is no solution to $I_i$, then go to 8.
  \item $\qquad \qquad$ Compute the expected measure of $\mathcal{R}(I_i)$
  \item $\qquad \qquad$ Remove the constraint $v_i = x$ from $I_i$
  \item $\qquad$ \parbox[t]{10.9cm}{Let $x_i \in D$ be the value which maximises the
  expected measure of $\mathcal{R}(I_i)$ in the computations in
  4--8. Create a new instance, $I_{i+1}$, which is identical to $I_i$
  except for the addition of the constraint $v_i = x_i$.}
  \item Return the unique solution to $I_{|V|+1}$.
\end{enumerate}
\caption{The algorithm in Theorem~\ref{thmaffinapx}} \label{fig:alg-affineapx}
\end{figure}

We claim that the following loop invariant holds in the algorithm:
before line 4 is executed it is always the case that the expected
measure of $\mathcal{R}(I_i)$ is at least $E$.

We first prove the correctness of the algorithm assuming that the loop
invariant holds. From the loop invariant it follows that the expected
measure of $\mathcal{R}(I_{|V|+1})$ is at least $E$. In $I_{|V|+1}$
there is, for each variable $v_i \in V$, a constraint of the form $v_i
= x_i$, therefore there is only one solution to $I_{|V|+1}$.  This
solution will be returned by the algorithm.

We now prove that the loop invariant holds. The first time line 4 is
reached the expected performance ratio of $\mathcal{R}(I_1)$ is at
least $E$, per the calculations above. Now assume that the loop
invariant holds in iteration $i = k \leq |V|$; we will prove that it
also holds in iteration $i = k+1$. Since the performance ratio of
$\mathcal{R}(I_k)$ is at least $E$, there must be some value $x \in D$
such that when $v_i$ is fixed to $x$, the performance ratio of
$\mathcal{R}(I_{k+1})$ is at least $E$. This element will be found by
the algorithm as it maximises the expected performance ratio
$\mathcal{R}(I_{k+1})$. Hence, the loop invariant holds for $i = k+1$.
\end{proof}

\subsection{Binary Commutative Idempotent Operation}
We now investigate the complexity of \prob{W-Max Sol$(\Gamma)$} for
maximal constraint languages $\Gamma$ satisfying $\langle \Gamma
\rangle = Inv(f)$ where $f$ is a binary commutative idempotent
operation.

Let $(F; +_F, -_F, \cdot_F, 1_F)$ be a finite field of prime order
$p$, where $+_F, -_F, \cdot_F$, and $1_F$ denotes addition,
subtraction, multiplication and multiplicative identity,
respectively (we refrain from defining a notation for multiplicative inverses, as we do not need it).
Furthermore, let $z_F$ be the unique element in $F$ such
that $z_F + z_F = 1_F$. Note that for $F = \mathbb{Z}_p$ we get $1_F =
1$ and $z_F = \frac{p+1}{2}$.

Let ${\mathcal A}$ denote the set of operations $f(x, y) = z_F \cdot_F
(x +_F y)$, where $F$ is a finite field of prime order $p = |D|$ and
$p > 2$. The proof will be partitioned into two main cases due to the
following result:

\begin{lemma}[\cite{BKJ01,S87}] \label{lem:notg}
If $Inv(f)$ is a maximal relational clone and $f$ is a binary idempotent operation, then either

\begin{enumerate}
\item
$Inv(f) = Inv(g)$ where $g \in {\cal A}$; or

\item
$B \in Inv(f)$ for some two-element $B \subseteq D$.
\end{enumerate}
\end{lemma}

The classification result is given in the next lemma together with a
proof outline. Full proofs concerning the case when $Inv(f) = Inv(g)$
and $g \in {\mathcal A}$ can be found in \S\ref{sec:case1}. In
\S\ref{sec:case2} we give a complete characterisation of the
complexity for the second case for domains $D$ such that $|D| \leq 4$.
Finally, in \S\ref{sec:bogdan} we extend the classification to general
domains under the assumption of a conjecture due to Szczepara
(Conjecture~\ref{conj:bogdan}).

\begin{lemma}\label{lem:bci}
  Let $f$ be a binary commutative idempotent operation on $D$ such 
  that $Inv(f)$ is a maximal relational clone, and let $\Gamma$ be a constraint 
  language such that $\langle \Gamma \rangle = Inv(f)$.

  \begin{itemize}

  \item If $Inv(f)=Inv(g)$ for some $g \in {\mathcal A}$, then {\scshape W-Max
      Sol}$(\Gamma)$ is \cc{APX}-complete;

  \item else if $|D| \leq 4$ and there exist $a,b \in D$ such that $a < b$ and $f(a,b)= a$, then
  let $a_*$ be the minimal such element (according to $<$), then

    \begin{itemize}
    \item {\scshape W-Max Sol}$(\Gamma)$ is \cc{poly-APX}-complete
      if $a_*=0$, and
    \item \cc{APX}-complete if $a_* > 0$.
    \end{itemize}

  \item Otherwise, if $|D| \leq 4$, then {\scshape W-Max Sol}$(\Gamma)$ is in \cc{PO}.
  \end{itemize}
\end{lemma}
\begin{proof}
If $Inv(f)=Inv(g)$ and $g \in {\mathcal A}$, then the result
follows from \S\ref{sec:case1}.

If there exist $a,b \in D$ such that $a < b$ and $f(a,b)= a$,
then we need to consider 
two cases depending on $a_*$. If $a_*=0$, then 
{\scshape W-Max Sol}$(\Gamma)$ is \cc{poly-APX}-hard by
Lemma~\ref{lemmahardness2} and
a member of \cc{poly-APX} by Lemma~\ref{lem-inpolyapx} since {\scshape Csp} is in
\cc{P}~\cite{BKJ01}. If $a_*>0$, then
{\scshape W-Max Sol}$(\Gamma)$ is \cc{APX}-complete by Lemma
\ref{binary:apxcomplete} in
\S\ref{sec:case2}.

Finally, if there do not exist any $a,b \in D$ such that $a < b$ and
$f(a,b)=a$, then $f$ acts as the $\max$ operation on every
two-element $B \subseteq D$ such that $B \in Inv(f)$.
Lemma~\ref{lem:twomax} shows that $f$ is a generalised $\max$ operation
in this case, and {\scshape W-Max Sol}$(\Gamma)$ is
in \cc{PO} by Theorem~\ref{thm-tractable-genmax}.
\end{proof}

\subsubsection{$f$ is contained in ${\cal A}$} \label{sec:case1}
We will now prove that {\scshape W-Max Sol}$(\Gamma)$ is
\cc{APX}-complete whenever $f \in {\cal A}$ and $\langle \Gamma \rangle = Inv(f)$.

\begin{lemma} \label{lem:A-apx-complete}
Let $f(x,y) = z_F \cdot_F (x +_F y)$, where $F$ is a finite field of
prime order $p = |D| > 2$ and $Inv(f)$ is a maximal relational clone.
Then, {\scshape W-Max Sol}$(\Gamma)$ is \cc{APX}-complete if $\langle \Gamma \rangle = Inv(f)$.
\end{lemma}
\begin{proof}
  We will give the proof for $F = \mathbb{Z}_p$ and after that we will
  argue that the proof can easily be adapted to the general case.

Let $q = \frac{p+1}{2}$ and $f$ be the function $f(x, y) = q(x + y)
\pmod{p}$. We will show that we can express $x - y + z$ through $f$.

Note that
  \begin{align} \label{eq:zero-sum}
  \sum_{i = 1}^{p-1} q^{i} = \frac{1 - q^p}{1-q} - 1 = 0 \pmod{p} .
  \end{align}
  (The second equality follows from Fermat's little theorem: $a^{p-1} =
  1 \pmod{p}$ for any prime $p$ and integer $a$ not divisible by $p$.)
  By using~\eqref{eq:zero-sum} and  Fermat's little theorem again, we get
  \begin{align} \label{eq:3sum}
  \sum_{i = 1}^{p-2} q^{i} = -1 \pmod{p} .
  \end{align}
  We can now express $x - y + z$ as follows:
  \begin{align} \notag
  &\underbrace{f(f(f(\ldots f(f(f(}_{p - 1 \textrm{ times}}x, z), y), y) \ldots), y), y) &= \\ \notag
  &q(q(q( \ldots q(q(q(x+z) + y) + y) + \ldots ) + y) + y)                       &= \\ \notag
  &q^{p-1} x + q^{p-1} z + \sum_{i = 1}^{p-2} q^{i} y                            &= \\ \notag
  &x - y + z \pmod{p}
  \end{align}
  where the final equality follows from~\eqref{eq:zero-sum},~\eqref{eq:3sum}
  and Fermat's little theorem.

  As any finite field $F$ of prime order is isomorphic to
  $\mathbb{Z}_p$, it is not hard to see that $x -_F y +_F z$ can be
  expressed through $f$ for any such field. Since $Inv(f)$
  is a maximal relational clone, $x -_F y +_F z$ can be expressed
  through $f$, and $x -_F y +_F z$ is not a projection, it follows that
  $Inv(f) = Inv(x -_F y +_F z)$. We now get containment in \cc{APX}
  from Theorem~\ref{thmaffinapx} and \cc{APX}-hardness from
  Theorem~\ref{th:affine-apx-hard}.
\end{proof}

\subsubsection{$f$ is not contained in ${\cal A}$} \label{sec:case2}
In the first part of this section we classify the complexity of
\prob{W-Max Sol}($\Gamma$) when $\langle \Gamma \rangle = Inv(f)$ for
all 2-semilattice operations $f$.
Recall that a 2-semilattice operation $f$ is an operation satisfying the conditions
$f(x,x)=x$, $f(x,y) = f(y,x)$, and $f(x,f(x,y)) = f(x,y)$.
It is noted in~\cite{Bulatov06}
that binary operations $f$ such that $Inv(f)$ is a maximal constraint
language on $|D| \leq 4$ are either 2-semilattices or otherwise
\prob{Csp}($\Gamma$) is \cc{NP}-complete. Hence, we get a
classification of the complexity of \prob{W-Max Sol}($\Gamma$) when
$\langle \Gamma \rangle = Inv(f)$ is a maximal constraint language
over $|D| \leq 4$ and $f$ is a binary operation.

The second result in this section is a complete complexity classification of \prob{W-Max Sol}($\Gamma$) for maximal constraint languages $\Gamma$, such that $\langle \Gamma \rangle = Inv(f)$ where $f$ is a binary operation, under the condition that Conjecture 131 from~\cite{S96} holds.

\begin{lemma}\label{lem:2semi1}
Let $f$ be a 2-semilattice operation on $D$ and $\langle \Gamma \rangle = Inv(f)$.
If there exist $a,b \in D$ such that $a < b$, $f(a, b) = a$, and $a^* > 0$ where $a^*$ is the
minimal element such that there is $b^*$ with $f(a^*, b^*) = a^*$,
then {\scshape W-Max Sol}$(\Gamma)$ is \cc{APX}-complete.
\end{lemma}
\begin{proof}
The \cc{APX}-hardness part is clear. What remains is to show that the problem is in \cc{APX}.
We can assume, without loss of generality, that $a = a^*$ and $b = b^*$.
We begin by proving that $U = D \setminus \{0\}$ is in $Inv(f)$.
Assume that $f(a,b)=0$ and $a,b > 0$, then $f(a,f(a,b)) = f(a,b) = 0$ and consequently $f(a,0) = 0$ contradicting the assumption that $a > 0$ was the minimal such element. Hence, $f(a, b) = 0$ if and only if $a = b = 0$. In particular $U$ is in $Inv(f)$.

We continue with the actual proof of the lemma.
Let $I=(V,D,C,w)$ be an arbitrary instance of
{\sc W-Max Sol}$(\Gamma)$. Define $V' \subseteq V$ such that
\[
V'=\{v \in V \mid \mbox{$S(v)=0$ for every solution $S$ of $I$}\}.
\]
We see that $V'$ can be computed in polynomial time: a variable $v$ is
in $V'$ if and only if the {\sc Csp} instance $(V,D, C \cup \{((v),
U)\})$ is not satisfiable.

Given two assignments $A,B:V \rightarrow D$, we define
the assignment $f(A,B)$ such that
$f(A,B)(v) = f(A(v),B(v))$. We note that if $A$ and $B$ are
solutions of $I$, then $f(A,B)$ is a solution to $I$, too: indeed,
arbitrarily choose one constraint $((x_1,\ldots,x_k), r) \in C$.
Then, $(A(x_1), \ldots, A(x_k)) \in r$ and
$(B(x_1), \ldots, B(x_k)) \in r$ which implies that
$(f(A(x_1),B(x_1)), \ldots, f(A(x_k),B(x_k))) \in r$, too.

Let $S_1,\ldots,S_m$ be an enumeration of all solutions 
of $I$ and define
\[
S^+ = f(S_1,f(S_2,f(S_3 \ldots f(S_{m-1},S_m) \ldots))).
\]
By the choice of $V'$ and the fact that $f(c,d)=0$ if and only if
$c=d=0$, we see that the solution $S^+$ has the following property:
$S^+(v)=0$ if and only if $v \in V'$. Let $p$ denote the second least
element in $D$, and note that $\opt(I) \geq \sum_{v \in V \setminus V'} w(v) p
= c$. Thus, by finding a solution with measure $\geq c$, we have
approximated $I$ within $(\max D)/p$ and {\sc W-Max Sol}$(\Gamma)$ is
in \cc{APX}.  To find such a solution, we consider the instance
$I'=(V,D,C',w)$, where $C' = C \cup \{((v), u) \mid v \in V \setminus
V'\}$.  This instance has feasible solutions (since $S^+$ is a
solution) and every solution has measure $\geq c$. Finally, a concrete
solution can be found in polynomial time by the result in~\cite{C04}.
\end{proof} 

\begin{lemma}
If $f$ is a 2-semilattice operation such that $f \not \in
\mathcal{A}$, $\Gamma$ is a maximal constraint language satisfying
$\langle \Gamma \rangle = Inv(f)$, and for all two-element $B \in
Inv(f)$ the operation $f$ acts as the $\max$ operation on $B$, then
{\scshape W-Max Sol}$(\Gamma)$ is in \cc{PO}.
\end{lemma}
\begin{proof}
What we will prove is that if $f$ acts as $\max$ on all two-element $B \in Inv(f)$, then  
$f$ is a generalised max operation and consequently {\scshape W-Max Sol}$(\Gamma)$ is in \cc{PO}. 

First note that if $a \neq b$ and $f(a,b) = a$, then by assumption $a>b$ and $f(a,b) > \min\{a,b\}$. 
Now, if $f(a,b) \neq a$, then $f(a,f(a,b)) = f(a,b)$ and by assumption $f$ is $\max$ on
$\{a,f(a,b)\}$. As a consequence of this we get $f(a,b) > \min\{a,b\}$.
Now, $f(a,b) > \min\{a,b\}$ for all $a \neq b$. Moreover, $f$ is idempotent, so $f$ is a generalised max operation and tractability follows from Theorem~\ref{thm-tractable-genmax}.
\end{proof}
 
We have now completely classified the complexity of {\scshape W-Max
Sol}$(\Gamma)$ for all constraint languages $\Gamma$ such that
$\langle \Gamma \rangle$ is maximal and $|D| \leq 4$.

\subsubsection{Complete Classification under a Conjecture} \label{sec:bogdan}
In this section we will prove that the validity of Conjecture 131
from~\cite{S96} implies a complete complexity classification of
{\scshape W-Max Sol}$(\Gamma)$ for all constraint languages $\Gamma$
such that $\langle \Gamma \rangle$ is maximal.
%(Lemmas \ref{binary:apxcomplete} and \ref{lem:twomax})
%are based on a conjecture by S... 
%This conjecture
%is known to be true for domains $D$ such that $|D| \leq 4$.
Given a binary operation $f$ on $D$, the {\em fixity} of $f$ is denoted
$\mathcal{F}(f)$ and is defined by
$$\mathcal{F}(f) = \{(x,y) \in D^2 \mid f(x,y) \in \{x,y\}\}.$$
The \emph{fixity-count} of $f$ is defined to be the cardinality of
$\mathcal{F}(f)$ and is denoted $|\mathcal{F}(f)|$.

\begin{conjecture}[\cite{S96}, Conjecture 131] \label{conj:bogdan}
If $Inv(f)$ is a maximal relational clone and $Inv(f') = Inv(f)$, then 
$|\mathcal{F}(f)| = |\mathcal{F}(f')|$.
\end{conjecture}

Although Conjecture~\ref{conj:bogdan} is not known to hold in the
general case, it has been verified for small domains. In particular,
it was shown in~\cite{S96} that for domains $D$ such that $|D| \leq 4$
the conjecture holds. Our proof builds on a construction that
facilitates the study of operation $f$---the details are collected in
Lemma~\ref{sameclone}. The underlying idea and the proof of
Lemma~\ref{sameclone} are inspired by Lemma~3 in \cite{BKJ01}.

Let $f$ be a binary operation on $D$ and define
operations $f_1,f_2,\ldots:D^2 \rightarrow D$ 
inductively:
$$f_1(x,y)=f(x,y)$$ 
$$f_{n+1}(x,y)=f(x,f_n(x,y)).$$

\begin{lemma} \label{sameclone}
Assume $f$ to be a binary commutative idempotent operation on $D$ such that 
$Inv(f)$ is a maximal relational clone and $Inv(f) \neq Inv(g)$ for every
$g \in {\cal A}$.
The following holds:
\begin{enumerate}
\item
$f|_{B}=f_n|_{B}$ for every $n \geq 1$ and every two-element $B \subseteq D$ in 
$Inv(f)$; and

\item
$Inv(f)=Inv(f_n)$, $n \geq 1$.
\end{enumerate}
\end{lemma}
\begin{proof}
1. Arbitrarily choose a two-element $\{a,b\}=B \subseteq D$ in $Inv(f)$. 
There are two possible binary commutative idempotent operations on $B$, 
namely $\max$ and $\min$. 
We assume without loss of generality that $f|_B=\max$ and prove the result 
by induction over $n$. 
Since $f_1=f$, the claim holds for
$n=1$. Assume it holds for $n=k$ and consider $f_{k+1}$.
We see that
$f_{k+1}(a,b)=f(a,f_k(a,b))$ and, by the induction hypothesis, $f_k(a,b)=\max(a,b)$.
Hence, $f_{k+1}(a,b)=\max(a,\max(a,b))=\max(a,b)$.

2. Obviously, $f_n \in Pol(Inv(f))$ and, thus, 
$Inv(f) \subseteq Inv(f_n) \subseteq R_D$. Since
$Inv(f) \neq Inv(g)$ for every $g \in {\cal A}$,
we know from Lemma~\ref{lem:notg} 
that there is some two-element $B \in Inv(f)$. By the proof
above, we also know that $f|_{B}=f_n|_{B}$ so $f_n|_{B}$ 
(and consequently $f_n$) is not a projection. 
Thus, $Inv(f_n) \neq R_D$, since $Inv(f') = R_D$ 
if and only if $f'$ is a projection. By the assumption that $Inv(f)$ is a 
maximal relational clone and the fact that 
$Inv(f) \subseteq Inv(f_n) \subsetneq R_D$, 
we can draw the conclusion that
$Inv(f) = Inv(f_n)$.
\end{proof}

We will now present some technical machinery that is needed for proving
Lemmas
\ref{binary:apxcomplete} and \ref{lem:twomax}.

\begin{lemma}[\cite{S96}, Lemma 28] \label{lem:bogdan}
Let $f$ be an idempotent binary operation and $n \in \mathbb{N}$.
Then, $\mathcal{F}(f) \subseteq \mathcal{F}(f_n)$.
\end{lemma}
\begin{proof}
Let $(x,y) \in \mathcal{F}(f)$. Then, either $f(x,y) = x$ or $f(x,y) = y$.
Now
$$ f(x,y) = x \Longrightarrow f_n(x,y) =
\underbrace{f(x,f(x,\dots,f(x,y)\dots ))}_{n \; times} =$$ 
$$\underbrace{f(x,f(x,\dots,f(x,x)\dots ))}_{n-1 \; times} = x
\Longrightarrow f_n(x,y) = x.$$
While
$$f(x,y) = y \Longrightarrow f_n(x,y) =
\underbrace{f(x,f(x,\dots,f(x,y)\dots ))}_{n \; times} =$$
$$\underbrace{f(x,f(x,\dots,f(x,y)\dots ))}_{n-1 \; times} \Longrightarrow
f_n(x,y) = y.$$
\end{proof}

Assuming that Conjecture~\ref{conj:bogdan} holds, we get the following corollary as a
consequence of Lemma~\ref{lem:bogdan}.
\begin{corollary}\label{cor:bogdan}
If $Inv(f)$ is maximal relational clone such that $f$ is commutative,
idempotent, and $(x,y) \in \mathcal{F}(f_k)$, (i.e. $f_k(x,y) \in
\{x,y\}$), then $\{x,y\} \in Inv(f)$.
\end{corollary}
\begin{proof}
By Lemma~\ref{lem:bogdan}, we have $\mathcal{F}(f) \subseteq
\mathcal{F}(f_k)$, and if 
Conjecture~\ref{conj:bogdan} holds, then $|\mathcal{F}(f)| = |\mathcal{F}(f_k)|$
which implies that
$\mathcal{F}(f) = \mathcal{F}(f_k)$. Now, if $f_k(x,y) \in \{x,y\}$, then
$f(x,y) \in \{x,y\}$ and by the commutativity
of $f$ we have $f(y,x) \in \{x,y\}$. Since $f$ is idempotent, it is
clear that $\{x,y\} \in Inv(f)$.
\end{proof}

We continue by introducing a digraph associated with the binary operation $f$. 
This digraph
enables us to make efficient use of Lemma~\ref{sameclone}. 
Given a binary operation $f:D^2 \rightarrow D$, we define $G_f=(V,E)$
such that $V=D \times D$ and $E=\{((a,b),(a,f(a,b))) \; | \; a,b \in D\}$.
We make the following observations about $G_f$: 

\begin{itemize}
\item[(1)]
an edge $((a,b),(a,c))$ implies that
$f(a,b)=c$;

\item[(2)]
every vertex has out-degree 1; and

\item[(3)]
there is no edge $((a,b),(c,d))$ with $a \neq c$.
\end{itemize}

We extract some more information about $G_f$ in the next three lemmas.

\begin{lemma} \label{gfacyclic}
The digraph $G_f$ contains no directed cycle.
\end{lemma}
\begin{proof}
Assume $G_f$ contains a directed cycle. Fact (3) allows us to assume
(without loss of generality) that the cycle is
$(0,1),(0,2),\ldots,(0,k),(0,1)$ for some $k \geq 2$.
Fact (1) tells us that $f(0,1)=2$, $f(0,2)=3$, $\ldots$, $f(0,k-1)=k$ and
$f(0,k)=1$.
Furthermore, one can see that $f_2(0,1)=3$, $f_2(0,2)=4$, $\ldots$, and
inductively $f_p(0,i)=i+p \; ({\rm mod} \; k)$. This implies that 
$f_k(0,1)=1+k \; ({\rm mod} \; k) = 1$.  
By Corollary~\ref{cor:bogdan}, $\{0,1\}$ is a subalgebra of $Inv(f)$ which 
contradicts the fact that $f(0,1)=2$.
\end{proof}

\begin{lemma} \label{gfpaths}
Every path in $G_f$ of length $n \geq |D|$ ends in a reflexive vertex,
i.e., $f_n(a,b) =c$ implies that $(a,c)$ is a reflexive vertex.
\end{lemma}
\begin{proof}
Assume that $G_f$ contains a path $P$ of length $n \geq |D|$.
This path can contain at most $|D|$ distinct vertices by fact (3).
Fact (2) together with the acyclicity of $G_f$ implies that at least
one vertex $v$ on $P$ is reflexive; by using fact (2) once again, 
we see that there exists exactly one
reflexive vertex on $P$ and it must be the last vertex.
\end{proof}

\begin{lemma} \label{reflexivevertex}
If $f_n(a,b)=c$, then $(a,c)$ is a reflexive vertex in $G_f$.
\end{lemma}
\begin{proof}
By Lemma~\ref{gfpaths}, every path in $G_f$ of length $n \geq |D|$ ends in a 
reflexive vertex. Hence, $(a,f_n(a,b)) = (a,c)$ is a reflexive vertex.
\end{proof}

\begin{lemma} \label{binary:apxcomplete}
Let $f$ be a binary commutative idempotent operation on $D$ such that
$\Gamma$ is a maximal constraint language satisfying $\langle \Gamma
\rangle = Inv(f)$.  If there exist $a,b \in D$ such that $a < b$, $f(a, b) = a$, and
$a^* > 0$ where $a^*$ is the minimal element such that there is $b^*$
with $f(a^*, b^*) = a^*$, then (assuming Conjecture~\ref{conj:bogdan})
{\scshape W-Max Sol}$(\Gamma)$ is \cc{APX}-complete.
\end{lemma}
\begin{proof}
We can assume, without loss of generality, that $a = a^*$ and $b = b^*$.
The \cc{APX}-hardness part follows from Lemma~\ref{lemmahardness}. What remains is to show that the problem 
is in \cc{APX}. We begin the proof by
proving that the unary relation $U=D \setminus \{0\}$ is a member of $Inv(f)$.
Consider the digraph $G_f$.
As we have already observed in Lemma~\ref{gfacyclic}, there are no cycles
$(a,b_1),\dots,(a,b_k),(a,b_1)$, $k \geq 2$, in $G_f$ and every path of
length $n \geq |D|$ ends in a reflexive vertex (by Lemma~\ref{gfpaths}). 
Obviously, no vertex $(a,0)$ ($a > 0$) in $G_f$ is reflexive since this
implies that 
$f(a,0)=0$ which is a contradiction.
In particular, there exists no path in $G_f$ of length $n \geq |D|$
starting in a
vertex $(a,b)$ ($a> 0$) and ending in a vertex $(a,0)$, since this implies
that $(a,0)$ is reflexive by Lemma~\ref{gfpaths}.

We can now conclude that  $f_n(a,b) > 0$ when $a > 0$: if $f_n(a,b)=0$, then $(a,0)$
is reflexive by Lemma~\ref{reflexivevertex} which would lead to a contradiction.
Hence,
$f_n(a,b) > 0$ whenever
$a,b \in D \setminus \{0\}=U$ so
$U$ is in $Inv(f_n)$, and by Lemma~\ref{sameclone}(2), $U$ is in $Inv(f)$, too.
We also note that
$U \in Inv(f)$ together with the assumption that $f(0,b)>0$ for all $b >0$ implies that
$f(c,d)=0$ if and only if $c=d=0$. The rest of the proof is identical to the
second part of the proof of Lemma~\ref{lem:2semi1}.
\end{proof}

\begin{lemma}\label{lem:twomax}
If $f$ is a binary commutative idempotent operation such that $f \not
\in \mathcal{A}$, $\Gamma$ is a maximal constraint language satisfying $\langle \Gamma \rangle = Inv(f)$, and for all
two-element $B \in Inv(f)$ the operation $f$ acts as the $\max$ operation on $B$,
then (assuming Conjecture~\ref{conj:bogdan}) {\scshape W-Max Sol}$(\Gamma)$ is in \cc{PO}.
\end{lemma}
\begin{proof}
What we will prove is that if $f \not\in {\cal A}$ and $f$ acts as $\max$ on all two-element $B \in Inv(f)$, then 
there exists a generalised max function $f'$ such that $Inv(f') = Inv(f)$ and, hence, 
{\scshape W-Max Sol}$(\Gamma)$ is in \cc{PO}. 

Recall that there are no cycles in $G_f$ and every path of length $n=|D|$
must end in a reflexive vertex by Lemmas~\ref{gfacyclic} and \ref{gfpaths}.
We also note that if $G_f$ contains a reflexive vertex $(a,c)$ with $a
\neq c$, then 
$f(a,c) = c$ 
and
$c > a$ since $f$ is assumed to act as the $\max$ operation on all
two-element 
$B \in Inv(f)$.

We now claim that $f_n$ is a generalised max operation. Arbitrarily choose
$a,b \in D$.
If $f_n(a,b) = c$ ($a\neq c$), then there is a path in $G_f$ from $(a,b)$ 
to a reflexive vertex
$(a,c)$, and $c>a$ as explained above.
If $f_n(a,b) = a$, then
$\{a,b\} \in Inv(f)$ by Corollary~\ref{cor:bogdan}.
Since $f(a,b)=\max(a,b)$, Lemma~\ref{sameclone}(1) implies that
$f_n(a,b)=\max(a,b)$.
Thus, $f_n$ is a generalised max-operation.
\end{proof}

\section{Homogeneous Constraint Languages} \label{secdich}
In this section, we will classify the complexity of \prob{Max Sol} when
the constraint language is \emph{homogeneous}. A constraint language
is called homogeneous if every {\em permutation relation} is contained in
the language.
\begin{definition}
  A relation $R$ is a \emph{permutation relation} if there is a
  permutation $\pi : D \rightarrow D$ such that
  \[
  R = \{ (x, \pi(x)) \mid x \in D \} .
  \]
\end{definition}
Let $Q$ denote the set of all permutation relations on $D$.  The main
result of this section is Theorem~\ref{th:hom-classification} which gives
a complete classification of the complexity of \prob{W-Max
  Sol$(\Gamma)$} when $Q \subseteq \Gamma$. The theorem provide the
exact borderlines between tractability, \cc{APX}-completeness,
\cc{poly-APX}-completeness, and \cc{NP}-hardness of finding a feasible
solution.

As a direct consequence of Theorem~\ref{th:hom-classification}, we get
that the class of injective relations is a maximal tractable class for
{\scshape W-Max Sol}$(\Gamma)$. That is, if we add a single relation
which is not an injective relation to the class of all injective
relations, then the problem is no longer in \cc{PO} (unless \cc{P} =
\cc{NP}).

Dalmau has completely classified the complexity of {\scshape
  Csp}$(\Gamma)$ when $\Gamma$ is a homogeneous constraint
language~\cite{Dalmau05}, and this classification relies heavily
on the structure of homogeneous algebras. An algebra is called
\emph{homogeneous} if and only if every permutation on its universe is
an automorphism of the algebra. We will not need a formal definition
of homogeneous algebras and refer the reader
to~\cite{M82,M64,Szendrei} for further information on their
properties. All homogeneous algebras
have been completely classified by Marczewski~\cite{M64} and
Marchenkov~\cite{M82}.

Our classification for the approximability of {\scshape W-Max
  Sol}$(\Gamma)$ when $\Gamma$ is a homogeneous constraint language
uses the same approach as in~\cite{Dalmau05}, namely, we exploit the
inclusion structure of homogeneous algebras (as proved
in~\cite{M82,M64}). By Theorem~\ref{relclone},
it is sufficient to consider constraint languages
$\Gamma$ that are relational clones. This is where the homogeneous
algebras comes in: the classification of homogeneous algebras gives us
a classification of all homogeneous relational clones, and in
particular their inclusion structure (lattice) under set inclusion.
We refer the reader to~\cite{Szendrei} for a deeper treatment
of homogeneous algebras and their inclusion structure.

\begin{figure}
\begin{center}
\includegraphics[scale=0.65]{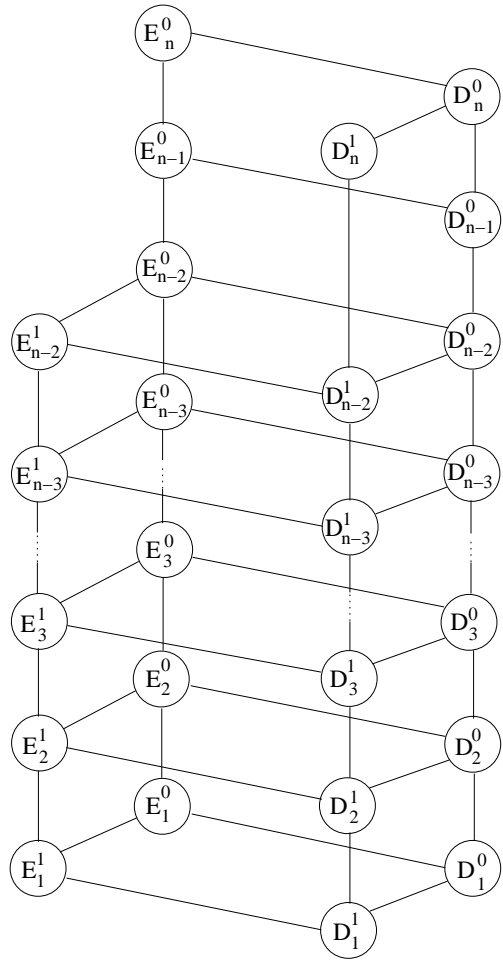}
\caption{Lattice of all homogeneous relational clones over domain size $n
\geq 5$.}
\label{hom5}
\end{center}
\end{figure}

\begin{figure}
\begin{center}
\includegraphics[scale=0.65]{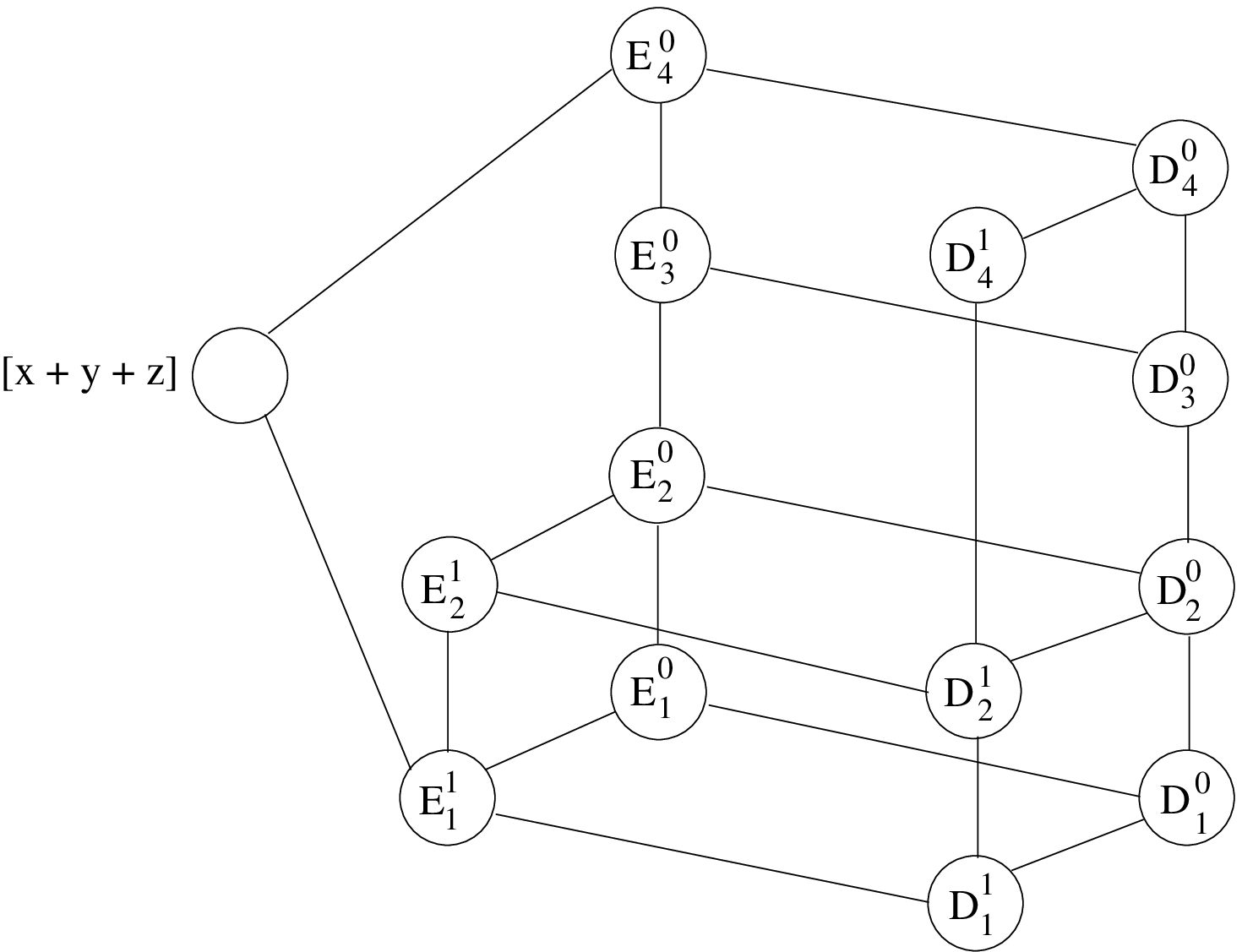}
\caption{Lattice of all homogeneous relational clones over domain size
$n=4$.}
\label{hom4}
\end{center}
\end{figure}

\begin{figure}
\begin{center}
\includegraphics[scale=0.65]{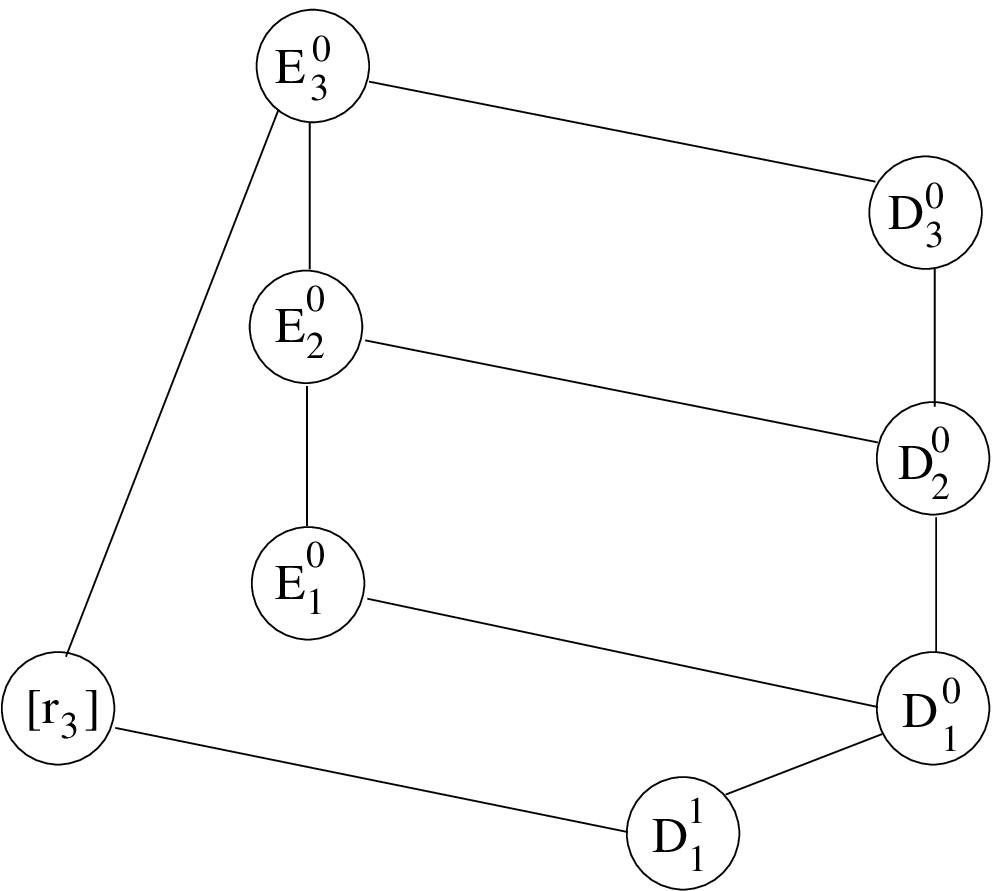}
\caption{Lattice of all homogeneous relational clones over domain size
$n=3$.}
\label{hom3}
\end{center}
\end{figure}

\begin{figure}
\begin{center}
\includegraphics[scale=0.65]{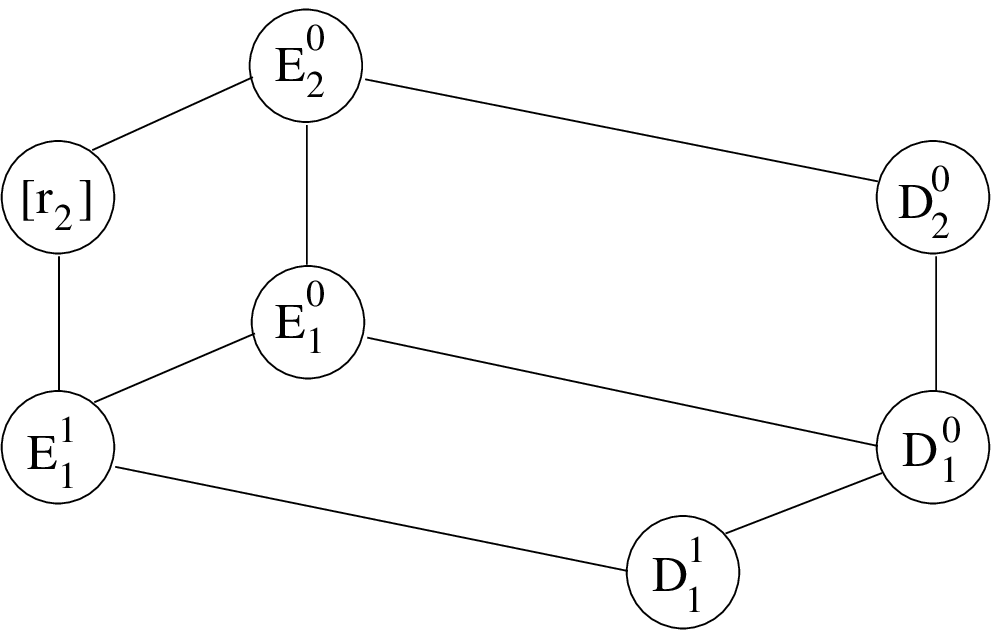}
\caption{Lattice of all homogeneous relational clones over domain size
$n=2$.}
\label{hom2}
\end{center}
\end{figure}

The lattice of all homogeneous relational clones on a domain $D$
having $n$ ($\geq 5$) elements is given in Figure~\ref{hom5}. The
lattices for the corresponding relational clones over smaller domains
(i.e., $2 \leq |D| \leq 4$) contains some exceptional relational clones
and
are presented separately in Figures~\ref{hom4}--\ref{hom2}. Note that
the corresponding lattices presented in~\cite{Dalmau05}
and~\cite{Szendrei} are the dual of ours, since they instead consider
the inclusion structure among the corresponding clones of operations
(but the two approaches are in fact equivalent as shown in~\cite[Satz 3.1.2]{PK79}).
To understand the lattices we first need some definitions.

Throughout this section $n$ denotes the size of the domain $D$, i.e., $n =
|D|$.
\begin{definition} \label{defoperations}
%We need to define the following operations.
\noindent
\begin{itemize}
\item
The {\em switching operation} $s$ is defined by
$$s(a,b,c) = \left\{ \begin{array}{ll}
c & \;\;\;\; \textrm{if $a = b$}, \\
b & \;\;\;\; \textrm{if $a = c$}, \\
a & \;\;\;\; \textrm{otherwise}. \end{array} \right.
$$
\item
The {\em discriminator operation} $t$ is defined by
$$t(a,b,c) = \left\{ \begin{array}{ll}
c & \;\;\;\; \textrm{if $a = b$}, \\
a & \;\;\;\; \textrm{otherwise}. \end{array} \right.
$$
\item
The {\em dual discriminator operation} $d$ is
defined by
$$d(a,b,c) = \left\{ \begin{array}{ll}
a & \;\;\;\; \textrm{if $a = b$}, \\
c & \;\;\;\; \textrm{otherwise}. \end{array} \right.
$$
\item
The $k$-ary {\em near projection operation} $l_k$ $(3 \leq k \leq n)$
defined by
$$l_k(a_1,\dots,a_k) = \left\{ \begin{array}{ll}
a_1 & \;\;\;\; \textrm{if $|\{a_1,\dots,a_k\}| < k$}, \\
a_k & \;\;\;\; \textrm{otherwise}. \end{array} \right.
$$

\item The $(n-1)$-ary operation $r_n$ defined by
\[
r_n(a_1,\dots,a_{n-1}) = \left\{
\begin{array}{ll}
a_1 & \;\;\;\; \textrm{if $|\{a_1,\dots,a_{n-1}\}| < n-1$}, \\
a_n & \;\;\;\; \textrm{otherwise.}
\end{array} \right.
\]
In the second case we have $\{a_n\} = D \setminus \{a_1, \ldots,
a_{n-1}\}$.

\item The $(n-1)$-ary operation $d_n$ (where $n \geq 4$) defined by
\[
d_n(a_1,\dots,a_{n-1}) = \left\{
\begin{array}{ll}
d(a_1,a_2,a_3) & \;\;\;\; \textrm{if $|\{a_1,\dots,a_{n-1}\}| < n-1$}, \\
a_n            & \;\;\;\; \textrm{otherwise.}
\end{array} \right.
\]
In the second case we have $\{a_n\} = D \setminus \{a_1, \dots,
a_{n-1}\}$.
\item
the operation $x+y+z$ where $(D,+,-)$ is a $4$-element group of exponent
$2$.
\end{itemize}
\end{definition}

The notation in the lattices in Figures~\ref{hom5}--\ref{hom2} is
explained below.
\begin{itemize}
\item
$D^0_1 = Inv(t)$;
\item
$D^0_i = Inv(\{d,l_{i+1}\})$ for $2 \leq i \leq n-1$;
\item
$D^0_n = Inv(d)$;
\item
$D^1_1 = Inv(\{t,r_{n}\})$;
\item
$D^1_i = Inv(\{d,l_{i+1},r_n\})$ for $2 \leq i \leq n-2$;
\item
$D^1_n = Inv(d_n)$;
\item
$E^0_1 = Inv(s)$;
\item
$E^0_i = Inv(l_{i+1})$ for $2 \leq i \leq n-1$;
\item
$E^0_n = R_D$, i.e., all finitary relations over $D$;
\item
$E^1_1 = Inv(\{s,r_n\})$ for $n \neq 3$;
\item
$E^1_i = Inv(\{l_{i+1},r_n\})$ for $2 \leq i \leq n-3$;
\item
$E^1_{n-2} = Inv(r_n)$ for $n \geq 4$.
\end{itemize}
Note that the relational clones described above depend on the size of
the domain $|D|=n$. Hence, $E^0_2$ in Figure~\ref{hom3} ($|D|=3$) is not
the same
relational clone as $E^0_2$ ($|D|=4$) in Figure~\ref{hom4}. Also note
that when we state our (in)approximability results by saying, for example
that,
{\scshape W-Max Sol}$(E^1_1)$ is \cc{APX}-complete, we mean that
{\scshape W-Max Sol}$(E^1_1)$ is \cc{APX}-complete for all sizes
of the domain where $E^1_1$ is defined (e.g., $E^1_1$ is not defined for
$n=3$). 
We always assume that $n = |D| \geq 2$.

We now state Dalmau's classification for the complexity of {\scshape
  Csp}$(\Gamma)$ for homogeneous constraint languages $\Gamma$.
\begin{theorem}[\cite{Dalmau05}]
\label{dalmauclass}
Let $\Gamma$ be a homogeneous constraint language. Then, {\scshape
  Csp}$(\Gamma)$ is in \cc{P} if $Pol(\Gamma)$ contains the dual
discriminator operation $d$, the switching operation $s$, or an affine
operation. Otherwise, {\scshape Csp}$(\Gamma)$ is \cc{NP}-complete.
\end{theorem}

We have the following corollary of Dalmau's classification.
\begin{corollary}
  Let $\Gamma$ be a homogeneous constraint language. Then, {\scshape
    W-Max Sol}$(\Gamma)$ is in \cc{poly-APX} if $Pol(\Gamma)$ contains
  the dual discriminator operation $d$, the switching operation $s$,
  or an affine operation. Otherwise, it is \cc{NP}-hard to find a
  feasible solution to {\scshape W-Max Sol}$(\Gamma)$.
\label{cordalmau}
\end{corollary}
\begin{proof}
  All dual discriminator operations, switching operations, and affine
  operations are idempotent (i.e., $f(x,x,x) = x$ for all $x \in D$).
  Hence, $\Gamma$ is invariant under a dual discriminator operation,
  switching operation, or an affine operation if and only if $\Gamma^c
  = \{\Gamma \cup \{\{(d_1)\}, \dots, \{(d_n)\}\}$ is invariant under the
  corresponding operation. Thus, it follows from
  Theorem~\ref{dalmauclass} that {\scshape Csp}$(\Gamma^c)$ is in
  \cc{P} if $Pol(\Gamma)$ contains the dual discriminator operation
  $d$, the switching operation $s$, or an affine operation. This
  together with Lemma~\ref{lem-inpolyapx} gives us that {\scshape
    W-Max Sol}$(\Gamma)$ is in \cc{poly-APX} if $Pol(\Gamma)$ contains
  the dual discriminator operation $d$, the switching operation $s$,
  or an affine operation.

  The \cc{NP}-hardness part follows immediately from
  Theorem~\ref{dalmauclass}.
\end{proof}

We begin by investigating the approximability of {\scshape W-Max
  Sol}$(\Gamma)$ for some particular homogeneous constraint languages
$\Gamma$.
\begin{lemma}
{\scshape W-Max Sol}$(D^0_n)$ is in \cc{APX} if $0 \notin D$ and in
\cc{poly-APX} otherwise.
\label{homcontainment}
\end{lemma}
\begin{proof}
  Remember that $D^0_n = Inv(d)$. Hence, membership in
  \cc{poly-APX} follows directly from Corollary~\ref{cordalmau}.  It
  is known from Dalmau's classification that {\scshape Csp}$(D^0_n)$ is
  in \cc{P}. Thus, by Proposition~\ref{propinapx}, it follows that
  {\scshape W-Max Sol}$(D^0_n)$ is in \cc{APX} when $0 \notin D$.
\end{proof}
\begin{lemma}
\label{lemd12}
{\scshape W-Max Sol}$(D^1_2)$ is \cc{APX}-complete if $0 \notin D$ and
\cc{poly-APX}-complete if $0 \in D$.
\end{lemma}
\begin{proof}
  Choose any $a,b \in D$ such that $a < b$. The relation
  $r=\{(a,a),(a,b),(b,a)\}$ is in $D^1_i = Inv(\{d,l_{i+1},r_n\})$ for
  $2 \leq i \leq n-2$. Hence, by Lemmas~\ref{lemmahardness}
  and~\ref{lemmahardness2}, it follows that {\scshape W-Max
    Sol}$(D^1_2)$ is \cc{APX}-hard if $0 \notin D$ and
  \cc{poly-APX}-hard if $0 \in D$.  This together with
  Lemma~\ref{homcontainment} and fact that $D^1_i \subseteq D^0_n$
  give us that {\scshape W-Max Sol}$(D^1_2)$ is \cc{APX}-complete if
  $0 \notin D$ and \cc{poly-APX}-complete if $0 \in D$.
\end{proof}

\begin{lemma} \label{leme12}
Finding a feasible solution to \prob{W-Max Sol}$(E^1_2)$ is \cc{NP}-hard.
\end{lemma}
\begin{proof}
Remember that $E^1_2 = Inv(\{l_{3},r_n\})$ when $n > 4$ and $E^1_2 =
Inv(r_n)$ when $n=4$, so it follows from Dalmau's classification
that {\scshape Csp}$(E^1_2)$ is \cc{NP}-complete.
\end{proof}

\begin{lemma}
{\scshape W-Max Sol}$(D^0_1)$ is in \cc{PO}.
\label{lemd01}
\end{lemma}
\begin{proof}
  It is well-known that $D^0_1 = Inv(t) = \langle I^D
  \rangle$; for instance, it is a direct consequence of
  Theorem~4.2 in \cite{Szendrei}.
  Hence, {\scshape W-Max Sol}$(D^0_1)$ is in \cc{PO} by the results in \S\ref{injectivetract}.
\end{proof}

\begin{lemma}
{\scshape W-Max Sol}$(E^0_1)$ is in \cc{APX}.
\label{leme01inapx}
\end{lemma}
\begin{proof}
  Remember that $E^0_1 = Inv(s)$.  Dalmau gives a polynomial-time
  algorithm for {\scshape Csp}$(Inv(s))$ in~\cite{Dalmau05} (he
  actually gives a polynomial-time algorithm for the more general class
  of para-primal problems). Dalmau's algorithm exploits in a clever way
  the internal structure of para-primal algebras to show that any
  instance $I$ of {\scshape Csp}$(Inv(s))$ can be split into
  independent subproblems $I_1, \dots, I_j$, such that
  \begin{itemize}
  \item the set of solutions is preserved (i.e., any solution to $I$
    is also a solution to each of the independent subproblems, and any
    solution to all of the independent subproblems is also a solution
    to $I$); and
  \item each $I_i$ ($1 \leq i \leq j$) is either an instance of
    {\scshape Csp}$(Inv(t))$ or an instance of {\scshape
      Csp}$(Inv(a))$, where $t$ is the discriminator operation and
    $a$ is an affine operation.
  \end{itemize}
  Hence, to show that {\scshape W-Max Sol}$(E^0_1)$ is in \cc{APX}, we
  first use Dalmau's algorithm to reduce the problem (in a solution
  preserving manner) to a set of independent {\scshape W-Max
    Sol}$(\Gamma)$ problems where $\Gamma$ is either invariant under
  an affine operation or the discriminator operation. We know from
  Lemma~\ref{lemd01} that {\scshape W-Max Sol}$(\Gamma)$ is in \cc{PO}
  when $\Gamma$ is invariant under the discriminator operation, and from
  Theorem~\ref{thmaffinapx} we know that {\scshape W-Max
    Sol}$(\Gamma)$ is in \cc{APX} when $\Gamma$ is invariant under an
  affine operation.  Since all the independent subproblems are in
  \cc{APX}, we get that the original {\scshape W-Max Sol}$(E^0_1)$
  problem is also in \cc{APX}.
\end{proof}

\begin{lemma}
\label{leme11}
{\scshape W-Max Sol}$(E^1_1)$ is \cc{APX}-complete.
\end{lemma}
\begin{proof}
  Remember that $E^1_1 = Inv(\{s,r_n\})$.  Note that $E^1_1 \subseteq
  E^0_1$ so membership in \cc{APX} follows from
  Lemma~\ref{leme01inapx}.

  For the hardness part, we begin by considering the general case when
  $n = |D| \geq 4$.  Choose an arbitrary two-element subset $\{a,b\}$
  of $D$ (without loss of generality assume that $a<b$) and let
  $(G,+,-)$ be the two element group on $G = \{a,b\}$ defined by $a +
  a=a$, $a+b = b+a = b$, and $b+b = a$.  Let $\Gamma$ be the set of
  all relations expressible as the set of solutions to equations over
  $(G,+,-)$. It is easy to realise that $\Gamma$ is invariant under $r_n$
  (since $r_n$ ($n \geq 4$) acts as a projection on $\{a,b\}$).
  Furthermore $\Gamma$ is invariant under $s$ since $s(x,y,z)$ acts as
  the affine operation $x+y+z$ on $\{a,b\}$. It is proved in
  Lemma~\ref{lemmaequationhardness} that {\scshape
    W-Max Sol Eqn}$(\mathbb{Z}_2, g)$ is \cc{APX}-hard so
  {\scshape W-Max Sol}$(\Gamma)$ is \cc{APX}-complete. For $n = |D| =
  3$ there is no relational clone of the type $E^1_1 =
  Inv(\{s,r_n\})$ so the only case that remains to be dealt with
  is the case where $n = |D| = 2$.

  Without loss of generality assume that $D = \{a,b\}$ where $a < b$.
  Again consider the group $(G,+,-)$ on $\{a,b\}$. Let $\Gamma =
  \{R_1, R_2\}$ where $R_1$ is the ($4$-ary) relation on $D$ which is
  the set of solutions to the equation $x_1 + x_2 + x_3 + x_4 = a$ and
  $R_2$ is the (binary) relation representing the set of solutions to
  the equation $y_1 + y_2 = b$.  Furthermore, let $\Gamma_{0,1} =
  \{R_1, R_2\}$ denote the special case where $a=0,b=1$ (i.e., $D=
  \{0,1\}$). It has been proved in~\cite[Lemma~6.9]{KSTW01} that
  {\scshape Max Sol}$(\Gamma_{0,1})$ is \cc{APX}-complete.  Let $I$ be
  an instance of {\scshape Max Sol}$(\Gamma_{0,1})$ containing $k$
  variables. It is easy to realise that if $I$ has a solution,
  then $\opt(I) \geq k/2$ (just note that the complement of any
  solution to $I$ is also a solution to $I$).  We give an $L$-reduction
  from {\scshape Max Sol}$(\Gamma_{0,1})$ to {\scshape Max
    Sol}$(\Gamma_{a,b})$.  Let $F(I)$ be the instance of {\scshape Max
    Sol}$(\Gamma_{a,b})$ where all occurrences of $0$ has been
  replaced by $a$ and all occurrences of $1$ has been replaced by $b$.
  Since $\opt(I) \geq k/2$, we get that $\opt(F(I)) \leq bk \leq 2b
  \cdot \opt(I)$ and $\beta = 2b$ is a valid parameter in the
  $L$-reduction. Let $s$ be an arbitrary solution to $F(I)$ and
  define $G(F(I),s)$ to be the corresponding solution to $I$ where $a$
  is replaced by $0$ and $b$ is replaced by $1$. Then,
  \[
  |m(I,G(F(I),s)) - \opt(I)| \leq \frac{1}{b-a}|m(F(I),s) - \opt(F(I))|
  \]
  and $\gamma = \frac{1}{b-a}$ is a valid parameter in the
  $L$-reduction. This completes the \cc{APX}-hardness proof for
  {\scshape Max Sol}$(\Gamma_{a,b})$. Now, the unary operation $r_2$
  acts as the non-identity permutation on $\{a,b\}$ (i.e., $r_2(a) =
  b$, and $r_2(b) = a$) so both $R_1$ and $R_2$ are invariant
  under $r_2$.  As we have already observed, $s$ acts as the affine
  operation on $\{a,b\}$ and $R_1$ and $R_2$ are invariant under $s$, too.
  Hence, $\Gamma_{a,b} \subseteq Inv(\{s,r_2\})$ which concludes the
  proof.
\end{proof}

We have now proved all the results needed to give a complete
classification for the approximability of {\scshape W-Max
  Sol}$(\Gamma)$ for all homogeneous relational clones $\Gamma$ over
domains $D$ of size at least $5$. In order to complete the
classification also for domains of size $2$, $3$ and $4$, we need to
consider some exceptional homogeneous relational clones. For domains of
size 4, we need to consider the homogeneous relational clone
$Inv(x+y+z)$ where $+$ is the operation of a $4$-element group
$(D,+,-)$ of exponent $2$ (i.e., a $4$-element group such that for all
$a \in D$, $a+a=e$ where $e$ is the identity element in $(D,+,-)$).
\begin{lemma}
\label{lemexp4}
Let $f(x, y, z) = x+y+z$ where $(D,+,-)$ is a group of exponent 2.
  Then, {\scshape W-Max Sol}$(Inv(f))$ is \cc{APX}-complete.
\end{lemma}
\begin{proof}
  The operation $x+y+x$ is the affine operation on $(D,+,-)$ (since
  $-y=y$ in $(D,+,-)$) and it follows directly from
Theorem~\ref{th:affine-apx-hard} and Theorem~\ref{thmaffinapx} that
  {\scshape W-Max Sol}$(Inv(x+y+z))$ is \cc{APX}-complete.
\end{proof}

For $3$-element domains, it remains to classify the approximability of
{\scshape W-Max Sol}$(Inv(r_3))$ where $r_3$ is the binary
operation defined as follows:
$$r(a_1,a_2) = \left\{ \begin{array}{ll}
a_1 & \;\;\;\; \textrm{if $a_1 = a_2$}, \\
a_3 & \;\;\;\; \textrm{where $\{a_3\} = D \setminus \{a_1,a_2\}$
otherwise}. \end{array} \right.
$$
\begin{lemma}
\label{lemr3}
{\scshape W-Max Sol}$(Inv(r_3))$ is \cc{APX}-complete.
\end{lemma}
\begin{proof}
  The operation $r_3(x,y)$ is actually an example of an operation of
  the type $\frac{p+1}{2}(x + y)$ (where $+$ is the operation of an
  Abelian group of order $|D| =p$) from
  Lemma~\ref{lem:A-apx-complete}. In our case $p=3$ and $r_3(x,y) = 2x
  + 2y$ where $+$ is the operation of the Abelian group $(D,+,-)$
  isomorphic to $\mathbb{Z}_3$. Hence, it follows from
  Lemma~\ref{lem:A-apx-complete} that {\scshape W-Max Sol}$(Inv(r_3))$
  is \cc{APX}-complete.
\end{proof}

For $3$-element domains we also need to classify the approximability of
{\scshape W-Max Sol}$(E^0_2)$ since hardness no longer
follows from the hardness of {\scshape W-Max Sol}$(E^1_2)$ (there is no
relational clone $E^1_2$ over $3$-element domains).
\begin{lemma} \label{leme02}
It is \cc{NP}-hard to find a feasible solution to {\scshape W-Max
Sol}$(E^0_2)$.
\end{lemma}
\begin{proof}
Immediate consequence of Theorem~\ref{dalmauclass}.
\end{proof}

Similarly, we also need to classify the approximability of {\scshape W-Max
Sol}$(D^0_2)$ since hardness no longer
follows from the hardness of {\scshape W-Max Sol}$(D^1_2)$ (there is no
relational clone $D^1_2$ over $3$-element domains).
\begin{lemma} \label{lemd02}
{\scshape W-Max Sol}$(D^0_2)$ is \cc{APX}-complete if $0 \notin D$ and
\cc{poly-APX}-complete if $0 \in D$.
\end{lemma}
\begin{proof}
  Remember that $D^0_2 = Inv(\{d,l_{3}\})$. It follows from
  the proof of Lemma \ref{lemd12} that {\scshape W-Max Sol}$(D^0_2)$
  is \cc{APX}-complete if $0 \notin D$ and \cc{poly-APX}-complete if
  $0 \in D$.
\end{proof}

For two-element domains $\{a,b\}$, we need to classify the unary
operation $r_2$ which acts as the non-identity permutation (i.e.,
$r_2(a) = b$ and $r_2(b) = a$).
\begin{lemma}
\label{lemhomlast}
Finding a feasible solution to \prob{W-Max Sol}$(Inv(r_2))$ is
\cc{NP}-hard.
\end{lemma}
\begin{proof}
Follows from Dalmau's classification.
\end{proof}

Finally, we are in the position to present the complete classification
for the approximability of all homogeneous constraint languages.
\begin{theorem} \label{th:hom-classification}
Let $\Gamma$ be a homogeneous constraint language.
\begin{enumerate}
\item If $\langle \Gamma \rangle \in \{E^i_j \mid i \in \{0,1\}, j
  \geq 2\}$ or $\langle \Gamma \rangle = Inv(r_2)$, then it is
  \cc{NP}-hard to find a feasible solution to {\scshape W-Max
    Sol}$(\Gamma)$;

\item else if, $0 \in D$ and $\langle \Gamma \rangle \in \{D^i_j \mid i
\in \{0,1\}, j \geq 2\}$,
  then {\scshape W-Max Sol}$(\Gamma)$ is \cc{poly-APX}-complete;
\item else if,
\begin{align}
&\langle \Gamma \rangle \in \{E^1_1, E^0_1, Inv(x+y+z), Inv(r_3)\} \quad
\textrm{or} \notag \\
&\langle \Gamma \rangle \in \{D^i_j \mid i \in \{0,1\}, j \geq 2\} \textrm{ and } 0 \notin D, \notag
\end{align}
then {\scshape W-Max Sol}$(\Gamma)$ is \cc{APX}-complete;

\item otherwise, $\langle \Gamma \rangle \in \{D^1_1, D^0_1\}$ and
  {\scshape W-Max Sol}$(\Gamma)$ is in \cc{PO}.
\end{enumerate}
\end{theorem}
\begin{proof}
  We know from Theorem~\ref{relclone} that it is sufficient to
  consider constraint languages that are relational clones.
  \begin{enumerate}
  \item It is proved in Lemmas \ref{leme12} and \ref{leme02} that it
  is \cc{NP}-hard to find a feasible solution to {\scshape W-Max
  Sol}$(E^1_2)$ and {\scshape W-Max Sol}$(E^0_2)$.  \cc{NP}-hardness
  of finding a feasible solution to {\scshape W-Max Sol}$(Inv(r_2))$
  was proved in Lemma~\ref{lemhomlast}. The result follows from the
  fact that $E^1_2 \subseteq E^1_j$ and $E^0_2 \subseteq E^0_j$ ($2
  \leq j \leq n$).

  \item Membership in \cc{poly-APX} is proved in
  Lemma~\ref{homcontainment} and \cc{poly-APX}-hardness of {\scshape
  W-Max Sol}$(D^1_2)$ and {\scshape W-Max Sol}$(D^0_2)$ when $0 \in D$
  is proved in Lemmas~\ref{lemd12} and \ref{lemd02}.  The result
  follows from the fact that $D^1_2 \subseteq D^1_j \subseteq D^0_n$
  and $D^0_2 \subseteq D^0_j \subseteq D^0_n$ ($2 \leq j \leq n$).

  \item It is proved in Lemmas~\ref{lemexp4} and~\ref{lemr3} that
  {\scshape W-Max Sol}$(Inv(x+y+z))$ and {\scshape W-Max
  Sol}$(Inv(r_3))$ are \cc{APX}-complete.  It is proved in
  Lemma~\ref{leme01inapx} that {\scshape W-Max Sol}$(E^0_1)$ is in
  \cc{APX}.  \cc{APX}-hardness for {\scshape W-Max Sol}$(E^1_1)$ is
  proved in Lemma~\ref{leme11}.  For $n \geq 4$ or $n=2$ we have that
  $E^1_1 \subseteq E^0_1$ and it follows that {\scshape W-Max
  Sol}$(E^0_1)$ and {\scshape W-Max Sol}$(E^1_1)$ are
  \cc{APX}-complete.  For $n=3$, there exists no $E^1_1$ so
  \cc{APX}-hardness for {\scshape W-Max Sol}$(E^0_1)$ must be proved
  separately. Since $E^0_1 = Inv(s)$ it is easy to see that the proof
  of Lemma~\ref{leme11} gives \cc{APX}-hardness for {\scshape W-Max
  Sol}$(E^0_1)$.
  
  Membership in \cc{APX} for {\scshape W-Max Sol}$(D^0_n)$ when $0
  \not \in D$ is the first part of
  Lemma~\ref{homcontainment}. \cc{APX}-hardness of {\scshape W-Max
  Sol}$(D^1_2)$ and {\scshape W-Max Sol}$(D^0_2)$ are proved in
  Lemmas~\ref{lemd12} and~\ref{lemd02}, respectively. Hence,
  \cc{APX}-completeness of {\scshape W-Max Sol}$(D^0_j)$ and {\scshape
  W-Max Sol}$(D^1_j)$ ($2 \leq j \leq n$) when $0 \notin D$ follows
  from the fact that $D^1_2 \subseteq D^1_j \subseteq D^0_n$, $D^0_2
  \subseteq D^0_j \subseteq D^0_n$ ($2 \leq j \leq n$).

  \item It is proved in Lemma \ref{lemd01} that {\scshape W-Max
  Sol}$(D^0_1)$ is in \cc{PO} and the result follows from the fact
  that $D^1_1 \subseteq D^0_1$.
  \end{enumerate}
\end{proof}

As a direct consequence of the preceding theorem, we get that the class
of injective relations is a maximal
tractable class for {\scshape W-Max Sol}$(\Gamma)$. That is, if we add
a single relation which is not an injective relation to the class of
all injective relations, then the problem is no longer in \cc{PO} (unless
\cc{P} = \cc{NP}).
\begin{corollary}
  Let $\Gamma_I^D$ be the class of injective relations
  and $R$ an arbitrary relation which
  is not in $\Gamma_I^D$.  Then, {\scshape W-Max
    Sol}\mbox{$(\Gamma_I^D \cup \{R\})$} is not in \cc{PO} (unless
  \cc{P} = \cc{NP}).
\end{corollary}
\begin{proof}
  We know from the proof of Lemma~\ref{lemd01} that $D^0_1 =
  Inv(t) = \Gamma_I^D = \langle I^D \rangle$. It follows from the
  lattices of homogeneous relational clones that either $E^0_1
  \subseteq \langle D^0_1 \cup \{R\} \rangle$ or $D^0_2 \subseteq \langle
D^0_1 \cup \{R\} \rangle$.
  We know from Lemma~\ref{leme11} and Lemma~\ref{lemd12} that both
  {\scshape W-Max Sol}$(E^0_1)$ and {\scshape W-Max Sol}$(D^0_2)$ are
  \cc{APX}-hard. Thus, {\scshape W-Max Sol}$(D^0_1 \cup \{R\})$ is
\cc{APX}-hard
  and it follows that {\scshape W-Max Sol}$(\Gamma_I^D \cup \{R\})$ is
  not in \cc{PO} (unless \cc{P} = \cc{NP}).
\end{proof}

\section{Conclusions} \label{sec:concl}
We view this article as a first step towards a better understanding of
the approximability of non-Boolean {\scshape W-Max Sol}. The ultimate
long-term goal for this research is, of course, to completely classify
the approximability for all finite constraint languages. However, we
expect this to be a hard problem since not even a complete
classification for the corresponding decision problem {\sc Csp} is
known.  A more manageable task would be to completely classify
{\scshape W-Max Sol} for constraint languages over small domains (say,
of size 3 or 4). For size 3, this has already been accomplished for
{\sc Csp}~\cite{CSP3} and {\sc Max Csp}~\cite{JKK06}.
Another obvious way to extend the results of this paper would be to complete the classification of maximal constraint languages over arbitrary finite domains, perhaps by proving Conjecture 131 from~\cite{S96}.

Our results combined with Khanna et al.'s~\cite{KSTW01} results for Boolean domains suggest the
following conjecture:

{\sc Conjecture.} For every constraint language $\Gamma$, one of
the following holds:

\begin{enumerate}
\item
{\scshape W-Max Sol}$(\Gamma)$ is in {\bf PO};

\item
{\scshape W-Max Sol}$(\Gamma)$ is {\bf APX}-complete;

\item
{\scshape W-Max Sol}$(\Gamma)$ {\bf poly-APX}-complete;

\item
it is {\bf NP}-hard to find a non-zero solution
to {\scshape W-Max Sol}$(\Gamma)$; or

\item
it is {\bf NP}-hard to find any solution to {\scshape W-Max Sol}$(\Gamma)$.

\end{enumerate}

If this conjecture is true, then there
does not exist
any constraint language $\Gamma_1$ such that \prob{W-Max Sol$(\Gamma_1)$} has
a polynomial-time approximation scheme ({\sc Ptas}) but \prob{W-Max Sol$(\Gamma_1)$}
is not in {\bf PO}. Natural such classes exist, however, if one restricts 
the way constraints are applied
to variables (instead of restricting the allowed constraint types).
{\sc Maximum Independent Set} (and, equivalently,
{\scshape Max Ones}$(\{(0,0),(1,0),(0,1)\})$) is one example: the unrestricted problem
is \cc{poly-APX}-complete and not approximable within $O(n^{1-\epsilon})$, $\epsilon > 0$
(unless {\bf P}={\bf NP})~\cite{Zuckerman06}, but the
problem restricted to planar instances admits a {\sc Ptas}~\cite{Baker:jacm94}.
One may ask several questions in connection with this: is there a constraint language
with the properties of $\Gamma_1$ above? For which constraint languages does
{\scshape W-Max Sol} admit a {\sc Ptas} on planar instances?
Or more generally: under which restrictions on variable scopes
does {\scshape W-Max Sol}$(\Gamma)$ admit a {\sc Ptas}?

It is interesting to note that the applicability of the algebraic
approach to {\scshape W-Max Sol} demonstrated in this article also holds
for the corresponding minimisation problem {\scshape W-Min Sol}, that
is, Theorem~\ref{relclone} still holds. The question whether the
algebraic approach can shed some new light on the intriguing
approximability of minimisation problems (as manifested, e.g.,
in~\cite{KSTW01}) is an interesting open question.

As mentioned in the introduction, it is known from~\cite{KSTW01} that
the approximability of the weighted and unweighted versions of
{\scshape (W)-Max Sol} coincide for all Boolean constraint languages.
We remark that the same result holds for all constraint languages
considered in this article. This can be readily verified by observing
that the $AP$-reduction from {\scshape W-Max Ones} to {\scshape Max
Ones} in the proof of Lemma 3.11 in~\cite{KSTW01} easily generalises
to arbitrary finite domains. Hence, we get an $AP$-reduction from
{\scshape W-Max Sol} to {\scshape Max Sol}. Furthermore, our
tractability proofs are given for the weighted version of the problem.
In general, it is still an open problem if tractability (i.e.,
membership in \cc{PO}) of \prob{Max Sol$(\Gamma)$} implies
tractability of \prob{W-Max Sol$(\Gamma)$} for every constraint
language $\Gamma$ (the $AP$-reduction used above do not give us this
result as $AP$-reductions do not, in general, preserve membership in
\cc{PO}).

\section*{Acknowledgements}
The authors would like to thank the anonymous referees for many
valuable comments and in particular for pointing out a serious flaw in
the (previous) proof of Lemma~\ref{lem:bci}. Peter Jonsson is
supported by the {\em Center for Industrial Information Technology}
(CENIIT) under grant 04.01 and the {\em Swedish Research Council} (VR)
under grants 621-2003-3421 and 2006-4532. Fredrik Kuivinen is
supported by VR under grant 2006-4532 and the {\em National Graduate
School in Computer Science} (CUGS).  Gustav Nordh is supported by CUGS
Sweden and Svensk-Franska Stiftelsen.

\bibliography{references}
\bibliographystyle{siam}
\end{document}